\documentclass[aps,prd,reprint,superscriptaddress, nofootinbib,
floatfix]{revtex4-2}
    
\usepackage{amsmath}
\usepackage{amssymb}
\usepackage{slashed}

\usepackage{amsxtra}
\usepackage{mathrsfs}
\usepackage{amsfonts}
\usepackage{arydshln}
\usepackage{dsfont}
\usepackage{dcolumn}
\usepackage{bm}
\usepackage{graphicx}
\usepackage{epstopdf}
\usepackage{txfonts}
\usepackage{braket}
\usepackage{comment}
\usepackage{blkarray}
\usepackage[dvipsnames]{xcolor}
\usepackage{color, colortbl}
\usepackage{cancel}
\usepackage{enumerate}

\usepackage[normalem]{ulem}

\usepackage[colorlinks=true,citecolor=blue,linkcolor=blue,urlcolor=blue]{hyperref}

\usepackage{orcidlink}

\usepackage{booktabs} 
\usepackage{siunitx} 
\usepackage{threeparttable} 
\usepackage{tabularx}
\usepackage{multirow}
\usepackage{diagbox}

\usepackage[utf8]{inputenc}

\usepackage{epsfig}
\usepackage{calc}
\usepackage[arrowdel]{physics}

\usepackage{lettrine}
\usepackage{Zallman}
\usepackage[capitalize]{cleveref}
\makeatletter
\AddToHook{cmd/appendix/before}{\def\cref@section@alias{appendix}\def\cref@subsection@alias{appendix}}
\makeatother
\crefname{section}{Sec.}{Secs.}

\makeatletter
\renewcommand{\@seccntformat}[1]{%
  \ifcsname prefix@#1\endcsname
    \csname prefix@#1\endcsname
  \else
    \csname the#1\endcsname\quad
  \fi}

\definecolor{niceblue}{rgb}{0.388235, 0.627451, 0.847059}
\definecolor{nicered}{rgb}{0.7,0.1,0.1}
\definecolor{nicegreen}{rgb}{0.1,0.5,0.1}
\definecolor{darkmagenta}{rgb}{0.55, 0, 0.55} 
\definecolor{persianblue}{rgb}{0.11, 0.22, 0.73}
\definecolor{LightCyan}{rgb}{0.88,1,1}

\newcommand{\pfrac}[2]{\left(\frac{#1}{#2}\right)}

\begin{document}

\rightline{DESY-26-086}

\title{\texorpdfstring{New energy-loss constraints on dark sectors \\ from 
deeply inelastic scattering with initial state radiation}{New energy-loss constraints on dark sectors from 
deeply inelastic scattering with initial state radiation}}

\newcommand{\ukyphys}{\affiliation{Department of Physics and Astronomy, University of Kentucky, Lexington, Kentucky, 40506-0055, USA}}

\author{Justin Cammarota\orcidlink{0000-0002-3076-201X}}
\ukyphys

\author{John Carlton\,\orcidlink{0000-0001-6483-5216}}
\ukyphys
\affiliation{Deutsches Elektronen-Synchrotron DESY, Notkestrasse 85, 22607 Hamburg, Germany}

\author{Susan~Gardner\orcidlink{0000-0002-6166-5546}}
\ukyphys

\date{\today}

\begin{abstract}
We employ the joint QED and QCD factorization of 
deeply inelastic, electron-proton scattering 
with generic initial state radiation to 
probe the possibility of  
exotic particle emission---i.e., of weakly coupled 
particles originating 
from a dark or hidden sector---through 
anomalous energy loss. We leverage this possibility 
through the consideration of phase-space-limited kinematic regions, for which the emission of 
an additional, undetected particle can particularly impact the associated cross section. 
In this first paper, as a
proof of principle, we focus on radiation from 
the incoming electron, 
considering  
the modification of 
the lepton distribution function
from the emission of particles, that could
have spin of up to $2$ and 
various, well-motivated electron couplings. 
We illustrate 
the sensitivity of
our approach through
the computation of the 
modified cross sections for the emission 
of MeV-GeV mass-scale, 
spin 0 
particles in 
kinematics chosen for
their sensitivity to initial state electron radiation and 
suitable to 
the forward-backward detection sensitivity of the 
ePIC detector at the Electron Ion Collider.      
\end{abstract}

\maketitle

\section{Introduction}

\lettrine{W}{e} propose 
a new search strategy for 
as-yet-unknown light particles 
at upcoming accelerator facilities, such as the Electron Ion Collider (EIC). 
Despite much astrophysical evidence for dark matter, 
its particle nature and nongravitational dynamics continue to be unclear~\cite{Chou:2022luk}. Moreover, 
the dynamical origin of the cosmic baryon 
asymmetry is not yet understood~\cite{Alarcon:2022ero}, and,  
with the discovery of gravitational waves~\cite{LIGO_2016PhRvX...6d1015A}, 
the prospects of identifying any footprints 
of quantum gravity beckon~\cite{deRham:2014zqa}. 
With these various motivations, searches for new
particles, nominally 
of a hidden or dark sector, 
have long 
been conducted in 
experiments at accelerators, either 
through bump hunt or displaced vertex searches. 
Here we offer a new possibility: constraining the 
emission of light, weakly coupled 
particles through anomalous energy loss, i.e., through 
the determination of the modification of the cross section that their emission would generate. 
To realize this new ``nonbump-hunt'' strategy, we employ the joint QED and QCD factorization of deeply  inelastic, 
electron-proton scattering~\cite{liu:2020rvc,liu2021new,cammarota2025factorizedqedqcdcontribution} with generic initial state radiation (ISR) to study the possibility of
exotic particle emission---of 
weakly coupled 
particles of varying mass and spin---that are undetected save for the energy loss they generate. 

The new, combined factorization approach 
reduces the uncertainties associated with the 
computed radiative corrections to electron-hadron 
scattering  
in  
traditional treatments, 
such as that of Mo and Tsai~\cite{Tsai:1961zz,RevModPhys.41.205,Tsai:1971qi}
or of the Dubna group~\cite{Bardin:1976qa,Bardin:1988by,Bardin:1989vz}, 
due to the parameter-dependent corrections they incur, which vary with the 
implementation~\cite{Badelek:1994uq}. Generally, 
electroweak radiative corrections to deeply
inelastic scattering (DIS), noting, e.g., \cite{Tsai:1961zz,RevModPhys.41.205,Tsai:1971qi,Bardin:1976qa,Bardin:1988by,Bardin:1989vz,Bardin:1989vz,Badelek:1994uq,Kripfganz:1990vm,Spiesberger:1994dm,Blumlein:2007kx}, can potentially dominate
the DIS cross section in regions
with more limited phase space, such as 
at large $y$ and $Q^2$. 
Under radiative corrections, the virtual 
photon can also develop a pinch singularity, 
which would 
need to be removed through a choice of kinematic cuts~\cite{cammarota2025factorizedqedqcdcontribution}\footnote{Or, this can be regulated theoretically by either 
limiting the energy of the radiated photon, 
or by giving the virtual photon an infinitesimal mass.}.
In the new 
 factorization approach, these limitations
 are addressed and removed at a formal level, 
 and the trade-off 
 is the introduction of nonperturbative functions associated with initial- and final-state leptons in the 
 scattering process to describe 
 possible radiation beyond QED. 
 These functions are not known exactly; 
 however, they can be determined from 
 the experimental data and their evaluation in some combination of 
 lattice  QCD with QED is an  
 open possibility\footnote{We note 
 \cite{patella_2017_qed_corrections} for a
 review of existing approaches and
 applications. More work must be done to show these new nonperturbative functions 
 can be calculated, possibly 
 in analogy to lattice QCD 
 computations of parton distribution functions (PDFs)~\cite{Lin:2025hka}.}. 
 In this way, the
 uncertainties in the 
 joint QED and QCD factorization method can be
 controlled, and here we study their impact
explicitly. A 
``QCD$\otimes$QED'' framework also 
exists~\cite{Cieri_2018_QED_QCD_qT_Z,Autieri_2023_QED_QCD_qT_WZ,buonocore_2024_qcdelectroweak_DY,autieri_2026_qT_qcdotimesqed,deFlorian:2025yar}, with different goals
from the combined framework we
use here, and we refer to~\cref{sec:QED+QCD}
for details. 

With the new QED and QCD 
framework in place, 
measurements of lepton-hadron DIS can be 
much more sensitive to 
energy-loss signatures, 
opening a new avenue to search for
physics beyond the Standard Model (SM).
The energy loss approach for detecting exotic particle 
emission that we espouse here can be applied broadly, 
though the range of possible 
particle dark matter masses is much broader still,  
ranging from as small as $10^{-22}\, \rm eV$, say, 
to $10^{15}$ eV or more~\cite{Chou:2022luk}. 
Here we focus on the discovery prospects associated
with the MeV-GeV mass scale, both because of
the particular capabilities 
of the upcoming EIC
and because that regime, at existing
facilities, has proven challenging 
experimentally, leaving large swathes of unexplored 
parameter space~\cite{Berlin:2018bsc,ema2025longlivedaxionlikeparticlestau}. 
Thus, the upcoming EIC, with a 
maximum
$e-p$ center of mass energy of 140 GeV~\cite{Abdul_Khalek_2022_short}, 
provides an ideal setting for non-SM 
searches in 
the MeV-GeV mass region. In this paper we focus on the DIS cross section, in 
different final states and kinematics, 
but the study of other 
observables may well 
prove fruitful. For example, 
energy loss studies can  
take advantage of the recent developments in energy-energy correlators~\cite{Moult:2025nhu,Guo:2025qnz}; however, such a study is beyond the scope of this work.

A broad range of beyond the SM (BSM) physics 
can be studied in this manner, though 
in this first paper 
we focus on the impact of new particles
that couple only 
to electrons.  
As long as the Lagrangian of the dark sector and its
interactions with electrons
is compatible with the SM, 
the contribution of 
the new particles to the DIS process 
can be 
computed 
in a manner similar to that of the SM radiative corrections in the QED and QCD factorization 
approach~\cite{liu:2020rvc,liu2021new,cammarota2025factorizedqedqcdcontribution}.
This work focuses on the corrections to the perturbatively calculable 
lepton distribution functions (LDFs) and fragmentation functions
(LFFs), since, as in QED, the non-SM particles are presumed
to be weakly coupled.
More generally,  
nonperturbative models for the LDFs and LFFs 
can potentially also be determined~\cite{liu:2020rvc,liu2021new,cammarota2025factorizedqedqcdcontribution},
though this would require the use of experimental data. 
Archival DIS datasets typically have 
some QED radiative corrections applied to them, and they  
would have to be removed in order for the data to be used in the
fitting process we envision. 
Here we use reference models in order
to determine the observables that can 
minimize 
their effects.

In this work, in order to leverage
the sensitivity of 
undetected, non-SM particle 
emission, we choose to study 
DIS with ISR from the incoming 
lepton line 
as our primary process, 
which is 
inspired by the use of ISR in $e^+e^-$ collisions 
at KLOE to probe the running of the fine-structure constant $\alpha_{\rm EM}$~\cite{Anastasi_2017_short}.  
In this paper, we also focus 
on the case in which the non-SM
particle 
is also emitted from the incoming  lepton, and this is possible because  final-state radiation has a smaller overall effect on the cross section. 
This plan 
also takes advantage of the far detection capabilities planned for the EIC~\cite{Abdul_Khalek_2022_short}, 
and this motivates why 
we use its specifics 
to guide our kinematic choices. 
Most ISR, at least in the kinematics we consider, 
is collinear, and this makes it difficult 
to detect in the primary ePIC detector. 
However, the far detectors along the direction of the lepton beamline 
allow for direct measurements of the photon's energy, which 
should 
help 
with background suppression and make
our desired candidate events more 
appreciable. 
This extra detection does require a slight modification to the QED and QCD factorization scheme 
we have mentioned; however, prior work has already 
shown how to address photoproduction in 
hadron collisions~\cite{BergerQiuPrompt} and the leading-order approximations 
of that paper can be applied here.

We conclude this section by 
outlining the content of those to follow. 
First, we briefly summarize
the standard QED and QCD factorization approach
and provide its essential formulas. 
We then turn, in the following section, 
to a discussion of the sorts of
non-SM particles of interest to us. 
If the new particles were massless, 
then Lorentz invariance would demand 
that their spin cannot be 3 or
higher---and
the massless spin 1 and 2 particles would 
couple via a conserved charge ($e$) 
and a universal graviton coupling, respectively~\cite{Weinberg_PhysRev.135.B1049}. 
Thus we reach beyond the photon and
the graviton by considering massive
particles. The spin 2 possibilities are especially 
interesting because of their connections 
to quantum gravity~\cite{Fierz:1939ix,deRham:2014zqa,Maggiore2007gwte.book.....M} as well. 
A broad range of candidates and candidate masses 
are possible, and in this paper, with our 
tilt to the capabilities of the EIC, we 
focus on candidates in the MeV-GeV mass range---and we develop detailed 
predictions for those with spin 0 only, 
for
simplicity. 
Then, in the following section,~\cref{sec:DIS+ISR}, 
the DIS with ISR process 
is described in more detail, along with more discussion
as to 
how to ensure its 
detection at the EIC. Subsequently, in~\cref{sec:results}, 
we show how particular kinematic choices 
that yield large values of 
the radiative 
corrections also leverage the largest impact on 
the DIS+ISR cross section from 
undetected, non-SM scalar particle emission. 
Finally, we conclude with a summary
and outlook.

\section{QED and QCD Factorization\label{sec:QED+QCD}}

DIS cross sections have long 
been calculated by separating contributions from process-dependent terms and process-independent terms through a controllable approximation method called (QCD) factorization~\cite{Collins:1985ue}. This requires two scales to distinguish between the  
hard and soft 
parts of the scattering process, or particularly that $Q^2 \equiv -(\ell - \ell^\prime)^2 \gg \Lambda^2_{\rm QCD}$ in the DIS
process $e(\ell)  h(P) \to e(\ell^\prime)  X$.
To describe the behavior of partons inside the hadron with 
their
longitudinal momenta, PDFs
are used to characterize the nonperturbative nature of the quarks and gluons~\cite{COLLINS1982445}.

Unobserved lepton radiation can drastically shift the cross section for lepton-hadron scattering processes, such as DIS, since 
the collision-induced lepton radiation changes the internal momentum transfer. The radiation can also alter the direction of the exchanged momentum, which means the 
expected 
Breit frame used  
to assess 
the
cross section (thus defining the PDFs) is different 
from 
the actual one the particles would 
possess.
The new QED and QCD factorization method~\cite{liu:2020rvc,liu2021new}, that we use 
here, 
puts QED and QCD radiation on an 
equal footing and removes this mismatch---and it also eliminates 
the pinch singularity coming from the exchanged 
photon, that we had noted earlier. 
The lepton analogs of the PDFs 
and parton fragmentation functions 
(FFs) for the quarks, the 
LDFs and LFFs, are defined for the incoming and outgoing leptons, respectively. This approach considers all QED radiation at the partonic scattering level and factorizes the collinear sensitive radiation into the LDF for leptons and PDFs for quarks (similarly for final state radiation into LFFs and FFs)~\cite{cammarota2025factorizedqedqcdcontribution}.

\begin{figure*} [hbt!]
  \centering
    \includegraphics[width=1\linewidth]{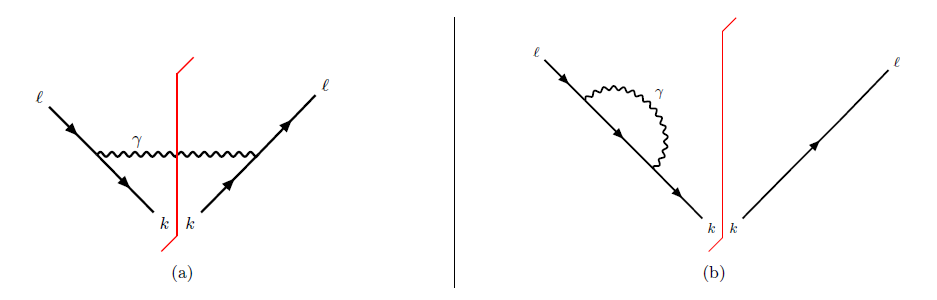}
    \caption{The  
    ${\cal O}(e^2)$ 
    cut graphs~\cite{Collins:2011zzd} for the 
    (a) real and (b) virtual contributions to the LDF 
    associated with $e(\ell) h(P) \to e(\ell^\prime)  X$, 
       where in b) the Hermitian conjugate must also be included.}
    \label{fig:RVLDF}
\end{figure*}

The LDFs and LFFs can contain 
hadronic
contributions, in principle, and thus 
nominally are nonperturbative objects. 
However, if those, 
such as could arise
from a radiated photon splitting to 
a $q\bar q$ pair that could also scatter with 
other quarks, are neglected, then 
the LDFs (and LFFs) become purely QED functions---and are perturbatively calculable in powers of $e$. The perturbative and nonperturbative
LDFs (and LFFs) are treated as independent functions~\cite{cammarota2025factorizedqedqcdcontribution}, and we refer to ~\cref{sec:results} for a discussion of their modeling 
and their impact on our analysis. 
We focus on the energy loss from ISR, so that the LDFs enter 
here, but the LFFs would be constructed similarly, and we 
refer to~\cite{liu:2020rvc,liu2021new,cammarota2025factorizedqedqcdcontribution} for
further discussion. With $\xi=k^+/\ell^+,$ as per the notation of~\cref{fig:RVLDF} and in light cone coordinates\footnote{We use the convention $v^{\pm}=(v^0\pm v^3)/\sqrt{2}$ for a generic vector 
$v$; here the $\hat{\mathbf{3}}$ direction points 
in the direction of the incoming lepton's momentum.} 
at leading order (LO),
$f_{e/e}^{(0)}(\xi)=\delta(\xi-1),$ which is intuitive as in this case 
there is no lepton-induced radiation. 
A vertical, broken line  
in~\cref{fig:RVLDF} 
indicates a 
``cut'' graph, where propagators crossing the cut are placed
on their mass shell, and we refer to \cite{Collins:2011zzd} for all 
details. In Next-to-Leading Order (NLO), 
$\alpha_{\rm EM}$ 
appears explicitly in the LDF, 
and this feature will 
become key when we turn to the analysis of non-SM particle emission.  The NLO contributions can be calculated from the Feynman diagrams 
in~\cref{fig:RVLDF}
in light cone gauge, i.e., 
$A^+=0$~\cite{Kogut:1969xa,Lepage_Brodsky_1980,COLLINS1982445}, 
as computed 
in~\cite{liu:2020rvc,liu2021new}, and the result in leading logarithmic order 
is  
\begin{align}\label{eq:ldf}
    f_{e/e}^{(1)}\left(\xi,\mu^2\right)&=\frac{\alpha_{\rm EM}}{2\pi}\left[\frac{1+\xi^2}{1-\xi}\ln\frac{\mu^2}{(1-\xi)^2 m_e^2}\right]_+ \,,
\end{align}
where the ``$+$'' prescription in the standard convention is 
\begin{align}
    [f(x)]_+&=f(x)-\delta(x-1)\int_0^1\dd{x} f(x) 
\end{align}
and $\mu$ is the factorization scale. 
The LDF combines with other 
functions, both hard and soft, 
to give the factorized DIS 
cross section, and this 
final quantity does not 
depend on $\mu$. This can be seen by enforcing
\begin{equation}
\frac{\dd{\sigma_{e(\ell) h(P)\to e(\ell') X}}}{\dd{\ln\mu^2}}=0 \,,
\end{equation}     
which yields evolution equations, similar to the DGLAP equations for the PDFs~\cite{Altarelli:1977zs,Gribov:1972ri}. For the 
LDF we have 
\begin{equation}\label{eq:evo}
    \frac{\dd{f_{e/e}\left(\xi,\mu^2\right)}}{\dd{\ln\mu^2}}=\int_{\xi}^1\frac{\dd{\xi'}}{\xi'}P_{e/e}\left(\frac{\xi}{\xi'},\alpha_{\rm EM} (\mu^2)\right) f_{e/e}\left(\xi',\mu^2\right)\,,
\end{equation}
which evolves the LDF from an initial factorization scale 
that combines to remove the overall $\mu$ dependence~\cite{liu:2020rvc,liu2021new,cammarota2025factorizedqedqcdcontribution}. Here $P_{e/e}$ is the evolution kernel, which is perturbatively calculable in a power expansion of $\alpha_{\rm EM}$ under QED alone~\cite{liu2021new}, while an expansion 
in terms of both $\alpha_{\rm EM}$ and $\alpha_S$ is possible  
if the factorization scale is sufficiently large~\cite{cammarota2025factorizedqedqcdcontribution}.
The factorization scale is conventionally chosen to be 
$\mu^2 = Q^2,$ but the transverse momentum of the outgoing lepton, $\mu^2= \ell_{T}^{\prime 2}=(1-y)Q^2,$ is also a valid choice~\cite{liu:2020rvc,liu2021new,cammarota2025factorizedqedqcdcontribution}. Because we are working in a collinear factorization approach, the $\ell_{T}^{\prime 2}$ must be large also, which is normally correlated to the large $Q^2$ conventionally chosen\footnote{Near the boundaries of the kinematic region, such as large $y$ and low $Q^2$ ($\sim 1$ GeV$^2$), this does not hold, see~\cite{cammarota2025factorizedqedqcdcontribution} for details.}. 

The factorized DIS cross section with only leading,  process-independent 
QED radiative contributions 
is given by
\begin{align}
&E'\frac{\dd[3]{\sigma^{{\rm RC(LO)}}_{e(\ell) h(P)\to e(\ell') X}}}{\dd[3]{\ell'}} 
\approx \frac{1}{2s}
\int_{\zeta_{\rm min}}^1 \frac{\dd{\zeta}}{\zeta^2}\, D_{e/e}\left(\zeta,\mu^2\right)
\nonumber \\
& \!\!
\times \int_{\xi_{\rm min}}^1 \frac{\dd{\xi}}{\xi}\, f_{e/e}\left(\xi,\mu^2\right)  \sum_{q}
\int_{x_{\rm min}}^1 \frac{\dd{x}}{x}\, f_{q/h}\left(x,\mu^2\right)
\widehat{H}^{(2,0)}_{eq\to eX}(\hat{s},\hat{t},\hat{u})\,,
\label{eq:lo-rc} 
\end{align}
and is labeled ``${\rm RC(LO)}$.''  
We have already noted the LDF
$f_{e/e} \left(\xi, \mu^2\right)$, and 
$D_{e/e} \left(\zeta,\mu^2\right)$ is the LFF, 
the function in the final state
analogous to it. 
Generally, the 
function $\widehat{H}^{(a,b)}_{eq\to eX}$ 
describes a  
contribution to the 
hard scattering,
partonic DIS process $eq\to eX$ 
in terms of powers of 
its coupling constants, namely,  
$\alpha_{\rm EM}^a$ and $\alpha_S^b$. Here 
 $\widehat{H}^{(2,0)}_{eq\to eX}$
 describes its first nonzero term
 in that power counting, thus giving 
 the LO contribution
 to the cross section. Generally, the hard function 
 $\widehat{H}^{(a,b)}_{eq\to eX}$ can contain 
 $\mu$ dependence from its renormalization, but 
 the leading term $\widehat{H}^{(2,0)}_{eq\to eX}$
 is finite and no such dependence appears. 

In 
this paper we work no higher than  
$\mathcal{O}\left(\alpha_{\rm EM}^2\right)$.
As we extend this framework to BSM physics, the cross section expression can be generalized to include other couplings,
such as would come from BSM-particle-fermion couplings. 
The relationship in \cref{eq:lo-rc} is only approximate, 
because 
the LDFs and LFFs are expanded to work
through NLO as 
we have described, though we will model and include
their nonperturbative contributions separately, as discussed 
in~\cref{sec:results}.
In contrast,  
the PDF is a fully nonperturbative
function 
determined from experimental data.
As this is a factorized cross section, there are also higher-order contributions at the presumed factorization scale, though these 
are suppressed as $\mathcal{O}(\Lambda_{\textrm{QCD}}^2/Q^2).$ 

In \cref{eq:lo-rc}, there are three integrations over the momentum fractions considered in the partonic scattering process. The first describes how 
the internal momentum $k^{\prime +}$ of 
the outgoing lepton produced in the scattering process is 
carried to that of the final-state observed lepton $\ell^{\prime +}$, and  $\zeta=\ell^{\prime}/k^{\prime} \approx 
\ell^{\prime +}/k^{\prime+}$. 
The second is the fraction of the initial-state lepton momentum entering the partonic interaction, that we have already defined. 
The last is the hadronic momentum fraction, following the standard definition for PDFs, $x=p^-/P^-,$ where $p$ is the momentum of the quark in the partonic 
scattering process in the CM frame, recalling $\ell$ is chosen
to be in the $+$ direction, 
and $P$ is that of its proton. 
The lower limits of integration are determined 
from the computed form of $\widehat{H}^{(2,0)}_{eq\to eX}$ 
and 
the Mandelstam variables $s=(\ell+P)^2$, $t=(\ell-\ell')^2$, and $u=(\ell'-P)^2$, with  
hatted variables denoting their partonic counterparts 
($\hat{s} = (k + p)^2$, $\hat{t}= (k - k^\prime)^2$, $\hat{u} = (k^\prime - p)^2$, where QED radiation from the quarks is implicitly absorbed in the PDFs),  
and noting 
 $x_B=Q^2/sy$ and $y=(P\cdot q)/(P\cdot \ell)$. The kinematic limits, 
 since the momentum fractions cannot be larger than 1, are explicitly given by~\cite{liu2021new,cammarota2025factorizedqedqcdcontribution}
\begin{align}
\zeta_{\rm min} &= - \frac{t+u}{s}
= 1- (1-x_B)\, y \, , \nonumber
\\
\xi_{\rm min} &= - \frac{u}{\zeta\, s + t}
=\frac{1-y}{\zeta-x_B y}\, , \nonumber
\\
x_{\rm min} &= - \frac{\xi\, t}{\xi\, \zeta\, s + u}
=\frac{\xi \, x_B \, y}{\xi\zeta +y -1}
\, , 
\label{eq:limits} 
\end{align} 
where we have assumed 
$s, u \gg m_e^2 + m_p^2$ and $t >> m_e^2$
throughout. 

For completeness, we note that 
other groups have worked to combine QCD and QED effects to hadronic interactions through 
so-called 
QCD$\otimes$QED frameworks, where they tend to focus on the resummation of logarithmic terms for the mixed contributions to the transverse momentum spectrum, especially for $W$ and $Z$ production in Drell-Yan at hadron colliders~\cite{Cieri_2018_QED_QCD_qT_Z,Autieri_2023_QED_QCD_qT_WZ,buonocore_2024_qcdelectroweak_DY,autieri_2026_qT_qcdotimesqed}. There has also been work that includes 
QED corrections to the Altarelli-Parisi splitting functions~\cite{deFlorian:2015ujt,deFlorian:2016gvk} and to realize 
mixed-order corrections to solutions to the DGLAP equations~\cite{deFlorian:2025yar}. 
The joint factorization method discussed here is different, as it 
places QED and QCD radiation in 
hadronic contributions on the 
same footing---and it is not limited to the transverse momentum resummation formalism.

\section{Extensions to BSM Physics\label{sec:BSM}}

We open with a survey of how new particles, from 
physics beyond the SM, may couple to 
SM fermions. Although our initial focus is on the case of BSM particles that couple to electrons, 
we wish to describe a sweep of possibilities in order to set the stage for later work. 
The observation of a cosmic baryon asymmetry, of dark matter, and of direct evidence for
the existence of gravitational waves---effects that are not of the SM of particle interactions---speak to the possibilities of new particles and forces, and the possible 
interactions they may have can depend on the dynamical function that they would possess. 
Since our new framework opens the possibility of probing new physics through energy loss in 
$e + p \to e+p + X$
scattering, 
we 
consider new particles of integer spin, even if
new particles 
in the MeV-GeV mass range, such 
as dark matter, could be fermions. 
These bosons may be portals to extended dark sectors, gauge bosons of new fundamental forces and as such may kinetically mix with existing force carriers in the SM. 

\subsection{Electron and dark-sector particle interactions}

We adopt an effective field theory approach to describe couplings between these new dark degrees of freedom and the SM,  
considering the broadest class of models. 
The primary consistency conditions we highlight are Lorentz invariance, gauge invariance, and unitarity. However, more subtle restrictions on theory parameter space also emerge from positivity bounds. 

Restricting ourselves to Lorentz invariant operators of the lowest dimension, the permitted interactions depend on the spin and parity of the new field such that for a field $\chi$, the interaction Lagrangian takes the form
\begin{equation}
    \mathcal{L_{\rm int}} \supset g_\chi\chi\mathcal{O}_e~.
\end{equation}
Here, $\mathcal{O}_e$ is the electron bilinear matching the Lorentz structure of the dark matter field $\chi$, and $g_\chi$ controls the coupling strength and production rate.

As we highlight in the introduction, the only viable massive Lorentz invariant fields we consider adding are spin 0, spin 1, and spin 2~\cite{Weinberg_PhysRev.135.B1049}. In the following subsections we review each of these classes of model and how viable they are as dark matter candidates. 
In addition to spin, parity provides a useful classification of possible mediators. For each spin assignment we consider both parity-even and parity-odd states whenever allowed. This leads to six representative scenarios for $J^{PC}$: scalar $(0^+)$, pseudoscalar $(0^-)$, vector $(1^-)$, axial-vector $(1^+)$, tensor $(2^+)$, and pseudotensor $(2^-)$. These cases provide a model-independent framework for studying missing-energy signatures in this work. We summarize the effective couplings in Table~\ref{tab:dmlagrangians}.

\begin{table}[ht]
    \centering
    \medskip
    \begin{tabular}{c c c}
        \toprule
        Model & $J^{PC}$ & Electron Interaction Lagrangian \\
        \midrule
        Scalar & $0^+$ & $g_{se} \phi \bar{\psi}_e\psi_e$ \\
        Pseudoscalar & $0^-$ & $g_{ae} a\bar{\psi}_e i\gamma_5\psi_e$\,, $g_{ae}  \partial_\mu a\,\bar{\psi}_e\gamma^\mu\gamma_5\psi_e /2 m_e$ \\
        Axial-vector & $1^+$ &  $g_{A'_{{\rm x}}e}\bar{\psi}_e\gamma^\mu\gamma_5\psi_e {A'_{\rm x}}_\mu$\\
        Vector & $1^-$ & $g_{A'e}\bar{\psi}_e\gamma^\mu\psi_e A^{\prime}_\mu$\\
        Tensor & $2^+$ & $g_{Te} h^{\mu\nu}\bar{\psi}_e\gamma_{(\mu}D_{\nu)}\psi_e$\\
        Pseudotensor & $2^-$ & $g'_{Te} h^{\mu\nu}\bar{\psi}_e\gamma_{(\mu}\gamma_5 D_{\nu)}\psi_e$\\
        \bottomrule
    \end{tabular}
    \caption{Dark sector candidates considered as possible modifications to the LDF and their electron interaction Lagrangians.
    }
    \label{tab:dmlagrangians}
\end{table}

A new scalar degree of freedom is perhaps the simplest additional field to incorporate into the 
SM. 
The lowest dimension couplings couple to the electron bilinears given by $\overline{\psi}_e\psi_e$ for a scalar field $\phi$ and $\overline{\psi}_e i\gamma^5\psi_e$ for a pseudoscalar $a$.  Such candidates arise naturally in models containing additional Higgs-sector singlets, dilatons, and axionlike particles~\cite{Silveira:1985rk,Marsh:2015xka, Cline:2013gha}, with important constraints, particularly on axionlike particles, 
coming from astrophysics and cosmology~\cite{Turner:1989vc,Marsh:2015xka}. 
While axions and axionlike particles are often thought of at lighter masses than the ones we consider in this work, there is nonetheless motivation to consider these heavier scalar candidates.

A massive spin 1 state $A^{\prime}_{\mu}$ may couple either to the conserved vector current $\overline{\psi}_e\gamma^\mu\psi_e$ or to the axial current $\overline{\psi}_e\gamma^\mu\gamma^5\psi_e$~\cite{Alexander:2016aln,Langacker:2008yv}. The vector case includes dark-photonlike scenarios arising from kinetic mixing with hypercharge, while the axial-vector interaction can emerge from a broken U(1) symmetry with chiral charge assignments. Since both currents are dimension three, the resulting interactions are renormalizable and constitute the dominant low-energy couplings of a massive vector mediator to electrons.

At leading order we expect a massive spin 2 field $h^{\mu\nu}$ to couple to the electron energy-momentum tensor $h^{\mu\nu}T_{\mu\nu}$.
This reduces to the tensor couplings shown in Table~\ref{tab:dmlagrangians}~\cite{Han:1998sg,donoghue2017epfllecturesgeneralrelativity, DAmico:2011eto}. We also include a parity-odd tensor interaction term as a phenomenological effective field theory benchmark with an additional $\gamma^5$ term in the coupling. However, no symmetric rank-2 tensor exists in the Standard Model to generate this coupling in analogy to $T_{\mu\nu}$, making the parity-even coupling better motivated for study.

Theories of spin 2 dark matter, and therefore of gravitons and gravitonlike particles, are less well defined in a  
field theory language. Without a complete description of quantum gravity, we instead rely on effective theories to describe the behavior of a hypothetical spin 2 field to address at least some of the pathologies. This can be achieved by considering a more concrete nonlinear theory such as dRGT massive gravity~\cite{deRham:2014zqa} or by breaking more fundamental symmetries such as Lorentz invariance~\cite{Rubakov:2004eb, Rubakov:2008nh, Blas:2024kps}. 

In this work we restrict ourselves to considering only Lorentz invariant fields as new degrees of freedom. Thus we employ the effective field theory of dRGT couplings of a massive spin 2 field (of mass $m$) with the lowest energy (high energy) cutoff scale given by $\Lambda_3=(m^2M_{\rm Pl})^{1/3}$. However, for an experiment performed on Earth we may instead use the redressed scale $\Lambda_*\sim10^7\Lambda_3$ (at the surface of the Earth), due to the Vainshtein mechanism changing the scale at which fluctuations become strongly coupled~\cite{deRham:2014zqa, Vainshtein:1972sx, Babichev:2013usa}. For a dark-sector particle mass on the order of MeV to GeV, we thus expect a cutoff of $\Lambda_*\sim (10^{10}-10^{13})~{\rm GeV}$, which ensures the validity of considering these couplings in our work. The undressed cutoff $\Lambda_3$ would also be sufficient at the energies we consider. Thus it may be possible to probe the validity of the Vainshtein mechanism through the effective spin 2 couplings we consider~\footnote{It is worth noting that as this spin 2 massive particle is not the graviton, the cutoff scale of the theory is not necessarily controlled by the usual Planck mass, instead this is a free parameter. However, $M_{\rm Pl}$ is a natural choice for new physics to enter in the gravity sector.}. 

Positivity bounds restrict self-interactions of massive tensor fields, but these align closely with the cutoff scale. Thus our effective theory remains valid for what we consider in this work~\cite{Alberte:2019xfh}. No pseudotensor analogy of dRGT currently exists. However, we assume such an effective theory would have a similar cutoff scale.

\subsection{Chiral basis}
The spin 0 and 1 interactions of Table \ref{tab:dmlagrangians}
can be captured in a compact way through the use of a chiral basis. 
Recalling that $L^a$ and $\ell^b$ are the left-handed doublets and
right-handed singlets of leptons in 
the SM, with $P_{R,L} = (1 \pm \gamma_5)$, we write, 
after ~\cite{ema2025longlivedaxionlikeparticlestau}, 
\begin{equation}
{\cal L}_{\rm int} = {\cal V}_\mu
[ g_{L{\cal V}}^{ab} \bar L^a \gamma^\mu P_L L^b 
+ g_{R{\cal V}}^{ab} \bar \ell^a \gamma^\mu P_R \ell^b ] \,,
\end{equation} 
where 
${\cal V}_\mu \in (A^\prime_\mu, {A_{\rm x}}^\prime_\mu, \partial_\mu \phi / 2m_e, \partial_\mu a / 2m_e)$ 
and $a,b$ are generational indices. Although
$(L^b)^T = (\nu^b, e^b)_L$, we do not consider
neutrino interactions in this paper. 
Since we are assuming parity invariance for 
simplicity, two sets of couplings, associated
with $P=+$ or $P=-$ mediators, are 
of interest, namely, 
\begin{eqnarray}
& g_{L{\cal V}}^{ab} = g_{R{\cal V}}^{ab} \equiv 
g_{{\cal V}_+}^{ab} \,, \nonumber \\
& g_{L{\cal V}}^{ab} = - g_{R{\cal V}}^{ab} \equiv
g_{{\cal V}_-}^{ab} \,
\end{eqnarray} 
with ${\cal V}_+ \in A_{\rm x}^\prime, s$ and 
${\cal V}_- \in A^\prime, a$. 
The terms in ${\cal L_{\rm int}}$ are each of 
mass dimension 5, but after integrating by 
parts and using the Dirac equation, we 
find the interactions for the 
scalar and pseudoscalar
candidates we use here. 
 Setting aside the possibility of 
charged lepton flavor violation, we let 
$a=b$ and consider electrons only henceforth---and we focus on the pseudoscalar case
in what follows\footnote{We note
that electromagnetic current conservation 
forbids a term of form 
$\partial_\mu \phi \bar \psi \gamma^\mu \psi$.}.  
We might suppose that the use of either
dimension 4 or 5 interaction terms
would be {\it equivalent}.
Yet there has been 
much discussion of this point in the literature, where we note~\cite{Berlin:2023ubt} for a detailed exposition, with the 
outcome that our simple derivation is only 
approximate---and that differences between
the two forms of the interaction can appear 
beyond leading order in the coupling constant. 
In the MeV-GeV region of interest to us, 
the possible
coupling constants need not be negligibly small, and 
experimental investigations of the differences 
may be possible. Thus we pause to flesh out the issues concretely. 

\begin{figure*}[hbt!]
   \centering
    \includegraphics[width=1\linewidth]{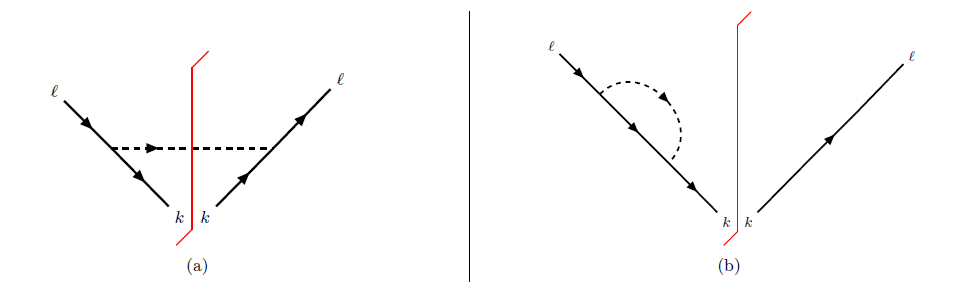}
    \caption{
    Cut graphs for the Feynman diagrams for the $\mathcal{O}(g_{se}^2)$  real 
    (a) and virtual (b) contributions to the LDF from a scalar, 
    where the Hermitian conjugate of (b) 
    also contributes.}
    \label{fig:RVSLDF}
\end{figure*}
The dimension 5 form of the 
$a-e$ interaction 
reveals that the physics should not 
change
under a constant shift of the fields 
$a$. Such a shift symmetry is not 
apparent in the noted 
dimension 4 term, however, and here we show how to 
connect the dimension
4 and 5 forms precisely. Moreover, the fermion
mass also plays a key role~\cite{Peccei:1977hh}, 
as it is a particular feature of the pseudoscalar case.
Writing $\tilde g_{ae} = g_{ae} / 2m_e $, 
the Lagrangian 
for an electron $\psi$ in the SM 
with an electron-axion 
interaction is 
\begin{equation}
{\cal L}_{{\rm f}\, a} \supset \bar\psi  (i \gamma^\mu D_\mu - m_e) \psi + \tilde g_{a e} \partial_\mu a  
\bar \psi \gamma^\mu \gamma_5 \psi \,,
\end{equation}
Here $D_\mu \equiv \partial_\mu + ie A_\mu $
if $e$ is the electron's physical charge. 
After redefining the fermion field
as per 
\begin{equation}
\psi \to e^{-i \tilde g_{ae} a \gamma_5 } \psi \,,
\end{equation}
we find, for the axionlike case, that~\cite{Adshead:2021ezw,Berlin:2023ubt} 
\begin{equation}
{\cal L}_{{\rm f}\, a} \supset \bar\psi 
\left( i \gamma^\mu D_\mu 
- m_e  e^{2i \tilde g_{ae} \gamma_5 a } \right) \psi 
- \frac{\alpha}{2\pi} \tilde g_{ae} a F_{\mu\nu} \tilde F^{\mu \nu} 
\,.
\end{equation}
The second term arises if $m_e \ne 0$
from the nonconservation of the axial vector current at the quantum 
level, due to the Adler-Bell-Jackiw anomaly~\cite{Adler:1969gk,Bell:1969ts,Adler:1969er}. 
Were we to ignore the impact of the
chiral anomaly, we see that the two interactions
would be equivalent to those in Table~\ref{tab:dmlagrangians} in linear
order of the coupling---but otherwise differ.
The term that arises from the chiral anomaly 
in the axion case requires special treatment in 
our framework, because although it can 
contribute as radiation from the electron
in the initial state, via $\gamma^\ast \to a \gamma$, 
this process is not collinear so that it cannot be ascribed to the perturbative 
LDF $f_{e/e} (\xi, \mu^2)$. This process, 
however, can contribute to the 
nonperturbative LDF. Thus we have shown 
that there is another role for BSM physics to play 
in the construction 
of this nonperturbative function, which we
reserve for future work. Now we turn 
to the assessment of the LDF under (collinear) 
BSM particle emission.

\subsection{Modifying the LDF with BSM physics}

Under the joint QED and QCD factorization framework, 
BSM 
particle emission
can easily be introduced through the modification of the LDF and LFF, using the interactions
listed in \cref{tab:dmlagrangians}.  
Here, the LDF can now include the emission of a scalar 
or an axionlike particle, as 
shown in~\cref{fig:RVSLDF}. 
Following~\cite{Collins:2011zzd}, which also analyzes
scalar emission from a fermion, we find for the real contribution to the emission 
of a scalar with mass $m_S$ that 
\begin{align}\label{eq:ldfscalar}
     f^{(1),\rm{s}}_R\left(\xi,\mu^2\right)&=\frac{g_{se}^2}{16\pi^2}(\xi-1)\left[\frac{\xi \left(m_S^2-4m_e^2\right)}{(1-\xi)^2 m_e^2+\xi m_S^2}\right.\nonumber\\
     &\left.-\ln\left(\frac{\mu^2}{(1-\xi)^2 m_e^2+\xi m_S^2}+1\right)\right]\,.
\end{align}
Following the same method, we can compute 
the LDF for the axionlike case (with mass $m_A$) using the dimension
4 interaction of Table~\ref{tab:dmlagrangians} 
to yield 
\begin{align}\label{eq:ldfaxion}
     f^{(1),\rm{a}}_R\left(\xi,\mu^2\right)&=\frac{g_{ae}^2}{16\pi^2}(\xi-1)\left[\frac{\xi m_A^2+2(1-\xi) \left(m_e^2\right)}{(1-\xi)^2 m_e^2+\xi m_A^2}\right.\nonumber\\
     &\left.-\ln\left(\frac{\mu^2}{(1-\xi)^2 m_e^2+\xi m_A^2}+1\right)\right]\,.
\end{align}
Thus we see that the two functions are distinct. 
However, if we employ 
the leading logarithm approximation, as we did in the QED case, 
then those differences disappear. One expects this 
approximation to give the numerically dominant 
contribution to the cross section, 
because it controls end point effects from the $\xi$ integration under the $+$ distribution. 
Finally, then, for the two cases, after combining the real and virtual contributions (which introduces the ``$+$'' prescription) and taking the leading logarithm approximation, we have 
\begin{align}\label{eq:ldfspin0}
  \!\!\!  f^{(1),0}\left(\xi,\mu^2\right)&=\frac{g_{ae}^2}{16\pi^2}\left[(1-\xi)\ln\left(\frac{\mu^2}{(1-\xi)^2 m_e^2+\xi m_A^2}\right)\right]_+\,.
\end{align}
Following a similar process 
for 
the LFF case, where here $k^+=(1/\zeta) \ell^{\prime +},$ the leading logarithm approximation gives 
\begin{align}
   \!\! D^{(1),0}\left(\zeta,\mu^2\right)&=\frac{g_{ae}^2}{16\pi^2}\left[(1-\zeta)\ln\left(\frac{\zeta^2\mu^2}{(1-\zeta)^2 m_e^2+\zeta m_A^2}\right)\right]_+ 
\end{align}
for both scalar and axionlike particle radiation. 
We think these degeneracies emerge as a 
consequence of the properties 
of the parity transformation on 
the light front. Since the 
spatial coordinate in the ${\hat{3}}$ direction of a 
vector $d$ is used to construct $d^{+, -}$, 
a suitable construction of the light-front 
parity operator on $d$ 
yields $d^\pm \to d^\pm$ and $d^{L,R} \to - d^{R,L}$, with 
$d^{L,R} \equiv d^1 \pm i d^2$~\cite{PhysRevD.73.036007}. 
Since our results do not depend on the transverse 
coordinates, we suppose that is why the 
LDF (and LFF) for each case turn out to 
be the same.

\section{DIS with ISR\label{sec:DIS+ISR}}

Motivated by the sensitivity of the DIS cross section to radiative corrections near the boundary of the kinematic phase space~\cite{liu:2020rvc,liu2021new,cammarota2025factorizedqedqcdcontribution}, we leverage its sensitivity 
to energy loss from BSM 
particle emission through the inclusion of SM 
ISR, such as that of a GeV-scale photon. 
Since the emitted BSM particles are undetected, 
experimental constraints on their mass are 
indirect. 

Here we find that by measuring GeV-scale ISR, we can produce
a kinematically restricted region that, in turn, is very 
sensitive to the emission of a leptophilic scalar and can 
also limit its mass. We note that the 
far-detection capacity of the EIC enables the realization of 
of this scenario. If the GeV-scale photon were not measured, 
we would lose sensitivity to scalar particle emission. Since we are measuring the emitted ISR photon, the limits imposed by knowing the photon energy provides an upper bound for the energy carried by the scalar, with an integration over the probabilities of all lower ones, but this provides an exact bound on the energy (and then directly on the mass if the scalar is at rest).

\subsection{SM DIS with ISR}\label{subDISwithISR}
As shown in \cite{BergerQiuPrompt}, photon production in a factorized process at leading order can be expressed in terms of the leading order partonic cross section (without photon production) convoluted with the corresponding distribution (or fragmentation) function where the detected photon's energy is used as a cutoff for the radiative phase space. In the case of initial state leptonic radiation in DIS considered here, this amounts to a change in the QED LDF convolution in~\cref{eq:lo-rc}, as a limit on $\xi.$ Instead of $\xi_{\rm min}$ in~\cref{eq:limits}, the minimum is found by considering the photon would have momentum $p_{\gamma}=(1-\xi)E_{\ell},$ which would correspond to $\xi_\gamma=1-p_\gamma/E_\ell,$ where $E_\ell$ is the energy of the incoming electron of momentum $\ell.$ Then the minimum in the LDF convolution is $\xi_{\rm ISR}=\max (\xi_\gamma,\xi_{\rm min}),$ as the kinematic constraints
we have noted 
would still need to be satisfied for this photon to be radiated from the incoming lepton. Thus, 
\begin{align}\label{eq:isr-lo}
&\!\!\!\!\!E'\frac{\dd[3]{\sigma^{{\rm RC(LO),ISR}}_{e(\ell) h(P)\to e(\ell')\gamma X}}}{\dd[3]{\ell'} } 
\approx \frac{1}{2s}
\displaystyle\int_{\zeta_{\rm min}}^1 \frac{\dd{\zeta}}{\zeta^2}\, D_{e/e}\left(\zeta,\mu^2\right)\nonumber \\
& \times \int_{\xi_{\rm ISR}}^1 \frac{\dd{\xi}}{\xi}\, f_{e/e}\left(\xi,\mu^2\right)  \nonumber\\
&\times\sum_{q}\int_{x_{\rm min}}^1 \frac{\dd{x}}{x}\, f_{q/h}\left(x,\mu^2\right)\widehat{H}^{(2,0)}_{eq\to eX}(\hat{s},\hat{t},\hat{u})\,,
\end{align}
where $\widehat{H}^{(2,0)}_{eq\to eX}(\hat{s},\hat{t},\hat{u})$ is 
determined by tree-level electron-quark scattering
\begin{align}\label{eq:dislo}
   \hat{H}^{(2,0)}_{eq\rightarrow eX}&=\frac{\alpha_{\rm EM} ^2 2^{2 \epsilon +2} \pi ^{\epsilon} e_Q^2 \left(v^2+(v-1)^2 (-\epsilon )+1\right)}{\hat{s} v (v-1)^2 \Gamma (1-\epsilon )}\nonumber\\
   &\times \left(\frac{1}{\hat{s} (1-v) v}\right)^{\epsilon }\delta (1-w)\,,
\end{align}
which is equivalent to the previously published result, 
though the $\epsilon$ dependence has been left explicit)~\cite{liu2021new,cammarota2025factorizedqedqcdcontribution}. Note that the dimensionless ratios
\begin{align}\label{eq:vwdef}
    v&=1+\frac{\hat{t}}{\hat{s}}\,,\nonumber\\
    w&=\frac{-\hat{u}}{\hat{s}+\hat{t}}
\end{align}
are used to simplify the expression. The LDFs and LFFs in~\cref{eq:isr-lo} are modeled using either the perturbative model as calculated above in~\cref{sec:QED+QCD} or using the nonperturbative model discussed below in~\cref{sec:results}. In this work, the LDF and LFF are taken to NLO, so for example, $f_{e/e}\left(\xi,\mu^2\right)=f^{(0)}_{e/e}\left(\xi,\mu^2\right)+f^{(1)}_{e/e}\left(\xi,\mu^2\right).$ However, the product of both the NLO LDF and LFF, $f^{(1)}_{e/e}\left(\xi,\mu^2\right)D^{(1)}_{e/e}\left(\xi,\mu^2\right),$ is dropped as it is suppressed by an additional power of $\alpha_{EM}.$ 

To evaluate the total section we recall that 
\begin{align}
    \frac{\dd[3]{\ell'}}{E'}&=-\frac{1}{2}\dd{y}\dd{Q^2}\dd{\phi}=\frac{Q^2}{2x_B}\dd{x_B}\dd{y}\dd{\phi}\,,
\end{align}
where $\phi$ is the angle between the plane defined by the incoming and outgoing lepton momenta and the plane defined by the hadron's spin and momentum~\cite{liu2021new}. 
Since we consider an unpolarized process, 
the integral over the 
angular dependence $d\phi$ simply gives
an overall factor of $2\pi$, implying that  
the cross section can be expressed as a differential of the  
usual, experimentally measured kinematic variables.

\begin{figure*}[hbt!]
    \centering
    \includegraphics[width=\linewidth]{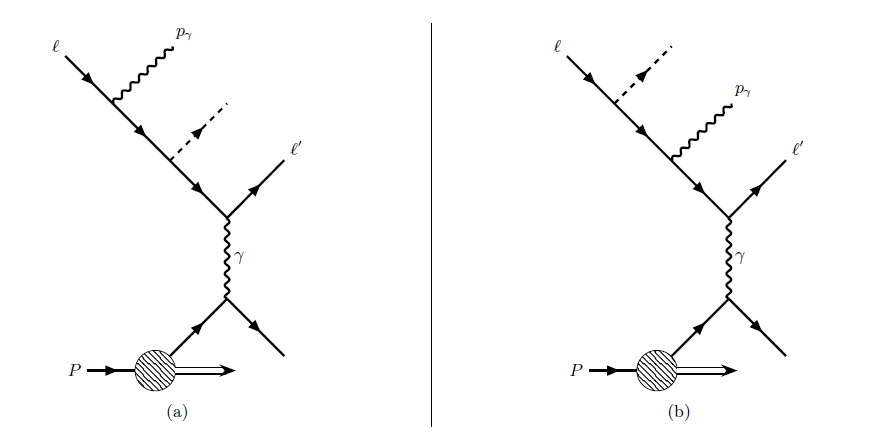}
    \caption{Feynman diagrams for DIS with combined QED and scalar ISR, with the scenario of the photon radiation occurring first in (a) or the scalar radiation occurring first in (b).}
    \label{fig:DISwISR_feyn}
\end{figure*}

The DIS with ISR process we discuss 
can be measured at the upcoming EIC. 
In the planned  
detector arrangement, the forward direction is taken to be in the direction of the hadron beam~\cite{Abdul_Khalek_2022_short}, which is opposite 
to the standard convention that we have also employed
in this paper. 
In order to ensure the detected photon comes from the initial-state lepton, it must be detected in the (now) backward region, along the path of the noninteracting electrons, as the LDF is dominated by collinear emission of the radiation. This also ensures the photon does not come from the quark or hadron, as backward emission {\it from the hadron} is kinematically unfavorable. The ability to distinguish between the sources of the photon radiation is a vital advantage to performing an experiment like this at the EIC, as their current plans include far detectors to observe exactly such radiation \cite{Abdul_Khalek_2022_short}.

\subsection{Inclusion of a calar}

Within this context 
it is important to understand how the signatures of BSM physics could be observed. In this paper we 
present a new technique of 
detecting the energy loss from BSM particle emission.
It is not a ``bump hunt'' at the resonant mass of the invisible particle, nor a displaced-vertex search, but, 
rather, a direct shift to the shape of the cross section as a
result of 
the different functions 
included 
in the LDF. The diagrams for the combined radiation are shown in~\cref{fig:DISwISR_feyn}, however, 
the leading order contribution $f^{(0)}$ for 
each particle contributes as well. 
The perturbative LDF for the cross section calculation in~\cref{eq:isr-lo} is modified by 
including the scalar LDF as calculated for a generic spin 0 particle in~\cref{sec:BSM}, and 
the factorized DIS cross section of~\cref{eq:lo-rc} is
also modified. Here we replace 
the integral over the LDF 
in~\cref{eq:lo-rc} as follows

 \begin{align}\label{eq:scalarLDF4ISR}
  &\int_{\xi_{\rm{ISR}}}^1\frac{\dd{\xi}}{\xi}f_{e/e}\left(\xi,\mu^2\right) \nonumber \\
    &\Rightarrow \int_{\xi_{\rm{ISR}}}^1\frac{\dd{\xi}}{\xi}f_{e/e}^{\rm{QED}}\left(\xi,\mu^2\right)\int_{\xi^{\prime}_{\rm{ISR}}}^1\frac{\dd{\xi^\prime}}{\xi^\prime}f_{e/e}^{\rm{S}}\left(\xi^\prime,\mu^2\right)\nonumber\\
    &\approx \int_{\xi_{\rm{ISR}}}^1\frac{\dd{\xi}}{\xi}f_{e/e}^{(0),\rm{QED}}\left(\xi,\mu^2\right)\int_{\xi_{\min}}^1\frac{\dd{\xi^\prime}}{\xi^\prime}f_{e/e}^{(0),\rm{S}}\left(\xi^\prime,\mu^2\right)\nonumber\\
    &+\int_{\xi_{\rm{ISR}}}^1\frac{\dd{\xi}}{\xi}f_{e/e}^{(1),\rm{QED}}\left(\xi,\mu^2\right)\int_{\xi_{\min}}^1\frac{\dd{\xi^\prime}}{\xi^\prime}f_{e/e}^{(0),\rm{S}}\left(\xi^\prime,\mu^2\right)\nonumber\\
    &+\int_{\xi_{\rm{ISR}}}^1\frac{\dd{\xi}}{\xi}f_{e/e}^{(0),\rm{QED}}\left(\xi,\mu^2\right)\int_{\xi_{\min}}^1\frac{\dd{\xi^\prime}}{\xi^\prime}f_{e/e}^{(1),\rm{S}}\left(\xi^\prime,\mu^2\right)\nonumber\\
    &+\int_{\xi_{\rm{ISR},1}}^1\frac{\dd{\xi}}{\xi}f_{e/e}^{(1),\rm{QED}}\left(\xi,\mu^2\right)\int_{\xi_{\rm{ISR},2}}^1\frac{\dd{\xi^\prime}}{\xi^\prime}f_{e/e}^{(1),\rm{S}}\left(\xi^\prime,\mu^2\right)\nonumber\\
    &+\int_{\xi^\prime_{\rm{ISR},1}}^1\frac{\dd{\xi^\prime}}{\xi^\prime}f_{e/e}^{(1),\rm{S}}\left(\xi^\prime,\mu^2\right)\int_{\xi^\prime_{\rm{ISR},2}}^1\frac{\dd{\xi}}{\xi}f_{e/e}^{(1),\rm{QED}}\left(\xi,\mu^2\right) \,,
\end{align}

where in the limits of integration shown  
we have specialized to the ISR case as per our discussion 
in Sec.~\ref{subDISwithISR}. 
Here $\xi'$ is the momentum fraction carried by the electron after the emission of the scalar,  and it is integrated through a convolution similar to that in 
$\xi.$ If both LDF contributions are taken in NLO, 
the order of the emission affects the kinematic limits, and 
its numerical subscript, either 
$\xi^{(')}_{\rm{ISR},1}$ or $\xi^{(')}_{\rm{ISR},2}$, 
reflects which one is emitted first. 
If the photon is emitted first, 
$\xi_{\rm{ISR},1}=\max (\xi_{\gamma},\xi_{\min})$ and $\xi_{\rm{ISR},2}=\hat{\xi}_{\min}=(1-\hat{y})\hat{s}/\left(\hat{s}-\hat{Q}^2\right),$ whereas if 
the scalar is emitted first, $\xi^\prime_{\rm{ISR},1}=\max (1-\xi_{\gamma},\xi_{\min})$ and $\xi^\prime_{\rm{ISR},2}=\max (\hat{\xi}_{\gamma},\hat{\xi}_{\min}),$ where $\hat{\xi}_{\gamma}=1-p_\gamma/(\xi^\prime E_\ell).$ 

The LFF is not expanded similarly, as in this work we
suppose the scalar is 
emitted from the initial state electron only. By taking the ratio of the DIS with ISR cross section for the various BSM physics models to the same DIS with ISR cross section for only SM behavior, significant deviations from unity can occur as a function of the
kinematics. Particularly, forming similar ratios of
the experimental cross section measurements and the
outcomes of different models, be it the SM or various BSM 
scenarios, 
can be used as evidence for exotic physics contributions---or can be potentially ruled out if the expected deviation 
does not occur.  
This requires a precise understanding of the associated 
LDFs and LFFs, both within and beyond the SM. The 
nonperturbative LDFs and LFFs that can arise from 
QCD interactions from photon splitting, 
such as into a $q\bar{q}$ pair at nonperturbative scales, 
where we note \cite{QW:evo} for further discussion, 
or simply from 
noncollinear emission effects, which can include BSM 
physics, should be determined from 
fitting the lepton functions to DIS experimental data.
As shown in~\cite{QW:evo}, 
the evolution kernels we used in \cref{eq:evo} 
address the broader class of splitting functions possible 
under the full QED and QCD framework. 
When BSM physics is included, 
new terms appear in 
the evolution kernels, though such effects
turn out to be negligible for the range of
empirically allowed scaling couplings.
We regard the nonperturbative LDFs as potentially reflective of the possibility of new physics nonetheless, because they should ultimately 
be fit to the experimental data---and in which 
the new scalar may appear. Given this setting, 
we advocate for new-physics probes 
in which the modeled nonperturbative LDF modifies the computed
observable in a minimal way---but for which the changes due to the use of the perturbative LDF with scalar emission can be 
readily determined.

Currently, there have not been any determinations
of the nonperturbative LDFs and LFFs, due to 
a lack of suitable lepton scattering data. 
This is because nearly all the  
existing, published lepton scattering data 
have already had radiative corrections applied, 
usually through the methods described in~\cite{Badelek:1994uq,Bardin:1976qa,Bardin:1989vz,Blumlein:2007kx,Kripfganz:1990vm,RevModPhys.41.205,Spiesberger:1994dm}. However, this unfortunate situation is slowly changing. 
For example, a recent analysis of ZEUS data explicitly defines the radiative correction factor~\cite{ZEUS:2023zie}, making
it possible to reconstruct the ``deradiated'' 
(uncorrected) data. 
Also an effort has been completed in which 
the needed uncorrected 
data has been 
reconstructed 
from old ALEPH data~\cite{Electron-PositronAlliance:2025fhk}, and upcoming world-wide experimental programs can also 
report their new data without radiative corrections. Under the inclusion of BSM physics, the nonperturbative model 
could be further modified, and indeed can offer 
another avenue by which to differentiate models, however adjustments to the evolution kernels alone do not indicate 
important effects.  

\section{Results\label{sec:results}}
\subsection{Scalar emission model\label{subsec:scalar}}

The DIS with ISR cross section reported 
in \cref{eq:isr-lo} has been evaluated 
at a particular choice of EIC kinematics, 
specifically, $\sqrt{s}=140$ GeV, $E_\ell=18$ GeV, and $0.01<y<0.95$~\cite{Abdul_Khalek_2022_short}, using CTEQ 18 unpolarized PDFs~\cite{Hou:2019efy}, and using either a perturbative expression {\it or}
a nonperturbative 
model for the LDFs and LFFs. 
The perturbative expression is given in~\cref{eq:scalarLDF4ISR}.  
The nonperturbative model employs 
the ansatz 
\begin{equation}
    f(x,\mu_0^2)=\frac{x^\alpha (1-x)^\beta}{B(\alpha+1,\beta+1)} \,,
    \label{eq:nonpertAnsatz}
\end{equation}
where we note \cite{Han:2021kes} for discussion of the quark
and gluon content of the electron at high energies. 
Here we suppose~\cref{eq:nonpertAnsatz} is given at 
an input factorization scale of $\mu_0=m_c$, 
the mass of the charm quark, so that our expression must 
be evolved to the scale pertinent to our analysis. 
To do this,  we use  
DGLAP-equivalent evolution equations for the LDF and LFF \cite{liu:2020rvc,liu2021new}.
Since 
our ansatz carries 
no explicit $\mu^2$ dependence,  
it is introduced through its evolution, using~\cref{eq:evo}. In this work, $(\alpha,\beta)$ are chosen to be $(5,0.5)$ and $(50,0.125)$ to match the behavior described in \cite{cammarota2025factorizedqedqcdcontribution}.
In previous work, it has been shown that 
large corrections may arise as a result of 
the nonperturbative models of the LDFs and LFFs~\cite{cammarota2025factorizedqedqcdcontribution}. We avoid this issue in two ways. First, we will only be comparing cross sections calculated using the joint QED and QCD factorization approach, making 
the effect of the 
corrections due to the nonperturbative model smaller, 
because the radiative corrections appear 
in both the scalar and SM-only calculations. Second, we will take ratios between 
observables calculated at 
different kinematics 
to reduce the impact 
of the ansatz parametrization. These ratios will be explained further in what follows. 

\begin{figure}[hbt!]
    \centering
    \includegraphics[width=\linewidth]{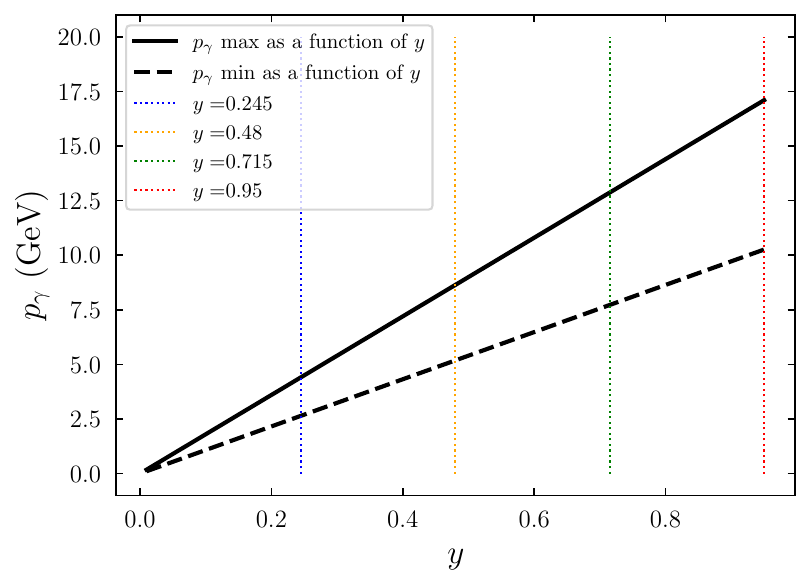}
    \caption{The variables $p_{\gamma,\max}$ and $p_{\gamma,\min}$ as a function of $y$ at specific EIC kinematics ($\sqrt{s}=140$ GeV and $E_\ell=18$ GeV).}
    \label{fig:pyvy}
\end{figure}

\begin{figure}[hbt!]
    \centering
    \includegraphics[width=\linewidth]{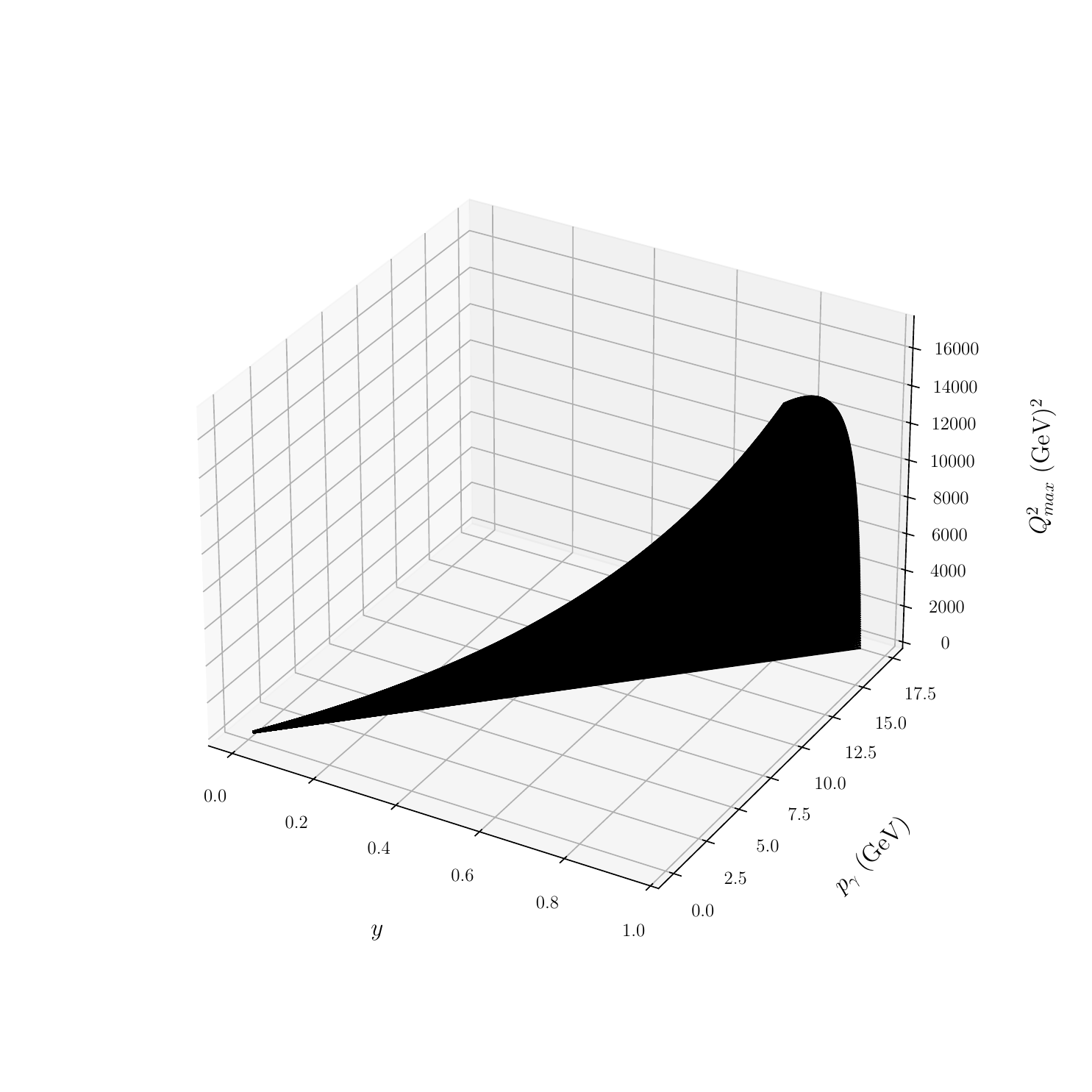}
    \caption{Kinematically allowable region for $Q^2_{\max}$ in terms of $p_\gamma$ and $y$ at specific EIC kinematics ($\sqrt{s}=140$ GeV and $E_\ell=18$ GeV).}
    \label{fig:q2vpvy}
\end{figure}

The kinematic limits for $p_\gamma$ and $Q^2$ are 
\begin{align}\label{eq:pqlimits}
    Q^2_{\max}&=\frac{s (E_\ell y-p_\gamma)}{E_\ell-p_\gamma}\nonumber\\
    p_{\gamma,\max}&=E_\ell \pfrac{s y-1}{s-1} \,.
\end{align}
The limits are found by enforcing $\xi_{\min}<\xi<1$, where the minimum comes from \cref{eq:limits} with $\zeta=1$, and $Q^2>1.$ The relation to $y$ is shown in \cref{fig:pyvy}, where the vertical dashed lines show the choices for $y$ used when comparing kinematic regions. A $p_\gamma$ minimum of $0.6 p_{\gamma,\max}$ 
is 
used to focus on the region where the perturbative and nonperturbative LDF (or LFF) show 
reasonable agreement, 
before 
additional 
modifications were studied, in order to 
minimize the differences resulting from 
the two descriptions. This limit also ensures the observed photon is hard for the process. The 
$Q^2_{\max}$ relationship is shown in \cref{fig:q2vpvy}, again with the minimum $p_\gamma$ cut we have noted. 

The specific kinematics within the EIC range  
are chosen to maximize the effects of the radiative corrections. As shown in~\cite{Badelek:1994uq}, the high $y$ region has the most impact from radiative corrections, as the corrections to the cross section can be 
of  
the same order of magnitude as the cross section itself. The work in~\cite{liu:2020rvc,liu2021new, cammarota2025factorizedqedqcdcontribution} extend this demonstration to the QED and QCD joint factorization method and similarly show the same kinematic region has a high impact from the LDF and LFF. By construction, the high $p_\gamma$ region implies a large shift from the effect 
of radiative corrections.

The mass and coupling choices for the scalar emission have been
made with \cite{ema2025longlivedaxionlikeparticlestau} in mind. 
We explore the mass range 
from $10^{-5}$ GeV to 
$1$ GeV, where the upper limit is 
constrained by the kinematics of the photon and scalar emission case, as the energy of the scalar must be less than the energy of the incoming lepton beam minus the energy of the radiated photon, 
and the remaining energy after both emissions must be large enough to
factorize DIS (i.e. $Q^2>1$ GeV$^2$). The coupling range explored was $10^{-9}\leq g_{se}\leq 10^{-2},$ which covers the range in \cite{ema2025longlivedaxionlikeparticlestau} for leptophilic scalars. All the observables considered below use $y=0.95,$ as this is the region with the largest possible radiative effects, but also the  
sharpest limit on possible scalar mass, which
emerges from allowing the largest possible $p_\gamma$. 
We also use the conventional choice of $\mu^2=Q^2$; 
however, we tested other choices for $\mu^2,$ such as $\mu^2=p_\gamma^2,$ 
and found 
no effects on the results. However, for low $y,$ $p_\gamma$ is not a reasonable choice for the factorization scale, as it
becomes too small, i.e., 
$p_\gamma^2 < 10$ GeV$^2$, thus potentially limiting 
the validity of the factorization 
approximation.

\subsection{Observables}

\begin{figure}[hbt!]
    \centering
    \includegraphics[width=\linewidth]{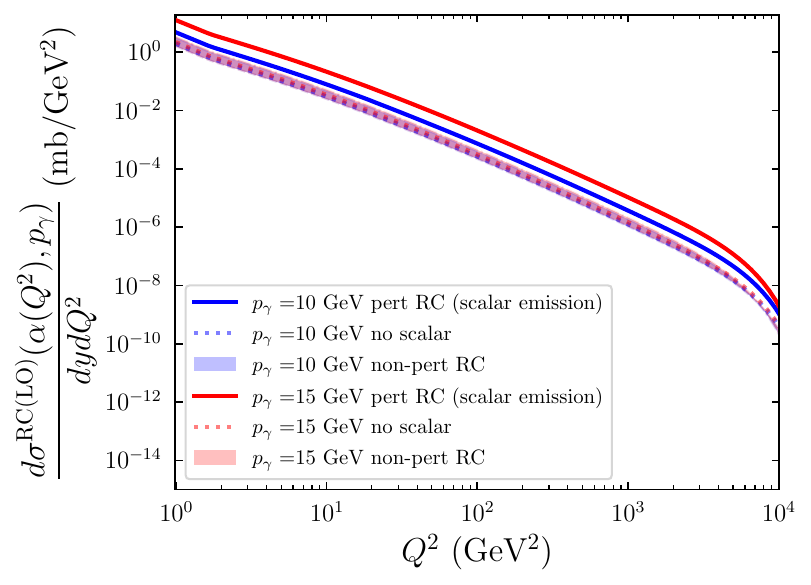}
    \caption{Cross section comparison for $m_S=0.9$ GeV and $g_{se}=3\times 10^{-6}$ at specific EIC kinematics ($\sqrt{s}=140$ GeV, $E_\ell=18$ GeV, and $y=0.95$).}
    \label{fig:xsec}
\end{figure}

Various DIS with ISR cross sections are shown in 
\cref{fig:xsec}. We show
the cross sections for DIS with ISR, that is, 
with an observed photon with energy $p_\gamma$ and unobserved scalar radiation, as well as 
DIS with ISR without any scalar emission. In 
what follows we consider various cross section 
ratios in order to display the effects of interest
more clearly. Our use of ratios should also 
reduce the sensitivity of our predictions to the
order of perturbation theory in which we work. As a first example, we consider 
\begin{align}\label{eq:obs1}
    \hat{\mathcal{O}}_1&=\frac{\sigma^{DIS,ISR (p_\gamma),s}}{\sigma^{DIS,ISR (p_\gamma)}}\,,
\end{align}    
where
\begin{align} 
    \sigma^{DIS,ISR (p_\gamma),s}&=\dfrac{\dd[2]{\sigma^\textrm{RC(LO),ISR,s}(\alpha(Q^2),p_\gamma,y)}}{\dd{y} \dd{Q^2}}\,, \nonumber\\
    \sigma^{DIS,ISR (p_\gamma)}&=\dfrac{\dd[2]{\sigma^\textrm{RC(LO),ISR}(\alpha(Q^2),p_\gamma,y)}}{\dd{y} \dd{Q^2}}\nonumber\,. 
\end{align}     

\begin{figure}[hbt!]
    \centering
    \includegraphics[width=\linewidth]{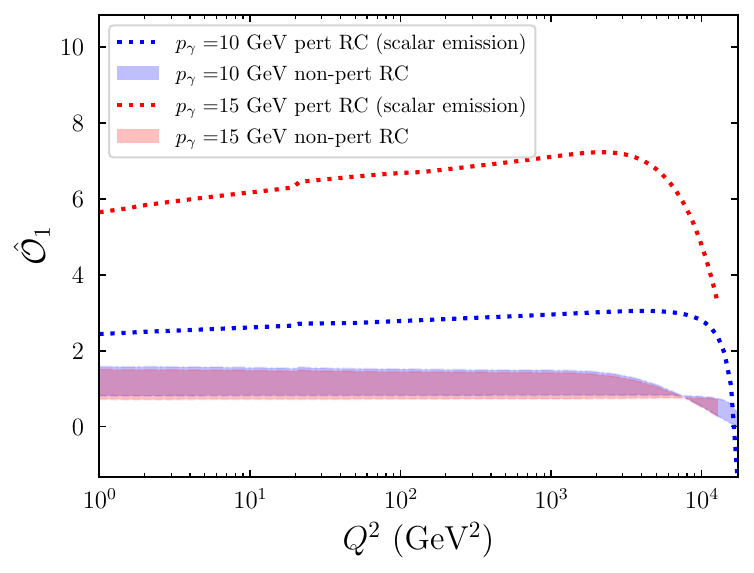}
    \caption{Observable 1 for $m_S=0.9$ GeV and $g_{se}=3\times 10^{-6}$ at specific EIC kinematics ($\sqrt{s}=140$ GeV, $E_\ell=18$ GeV, and $y=0.95$).}
    \label{fig:example3}
\end{figure}

This ratio, $\hat{\mathcal{O}}_1$ or observable 1, is shown 
in \cref{fig:example3}, and we see that 
the effects of the scalar emission can reach over five times the QED-only cross section. In all cases we have used the SM cross section computed with the perturbative LDF in the denominator and have 
included the running of $\alpha_{\rm EM}(Q^2)$. 
Here we use the
nonperturbative LDF shown in 
~\cref{eq:nonpertAnsatz} as a proxy for the additional possibility of
non-SM effects, such as the scalar emission model we assume here, 
in that soft function. The bands from the 
nonperturbative ansatz for the LDF 
appear to be well-separated from that of the 
perturbative LDF cases, but they 
nevertheless suggest the need to extract
those LDFs precisely to ensure that 
they do not introduce more uncertainty, 
which appears to be more
impactful in lower kinematic regions, such as in $y$ or $Q^2$.

Our analysis exists in advance of 
suitable experimental data. In the future, 
the EIC data, e.g., can be used to replace either model in the calculation of the ratios. 
If the data is assumed to follow the SM, 
then it can serve as the QED-only model in these ratios, and deviations from the SM 
could very well appear differently than in our plots,  
possibly yielding a ratio of
approximately unity  
if it follows the 
BSM model we have  noted.  
If, rather, we assume the data follow our 
scalar emission model, 
then the comparison should appear closer to our plots here, as our results 
show the deviations from the leading 
QED-only radiation, 
which is the dominant SM contribution.

There are further ratios we can consider. 
We can consider the ratio of the DIS
with ISR cross sections, with or without 
the unobserved scalar radiation, evaluated
in two different kinematic regions. 
In that case, one region can be chosen
so that it is  
less kinematically favorable to strong ISR. 
Thus the effects of the radiation can be observed also,  and we provide 
examples in what follows. Namely, we
consider ratios computed at the 
different values of $y$, as in the ratio $\hat{\mathcal{O}}_2$, 
observable 2, given by 
\begin{align}\label{eq:ratio1}
    \hat{\mathcal{O}}_2&=
    \frac{\sigma_{y2}}{\sigma_{y1}}
\end{align}
with either 
\begin{align} 
    \sigma_{y2}&=\dfrac{\dd[2]{\sigma^\textrm{RC(LO),ISR,s}(\alpha(Q^2),p_\gamma,y_2)}}{\dd{y} \dd{Q^2}}\nonumber\\
    \sigma_{y1}&=\dfrac{\dd[2]{\sigma^\textrm{RC(LO),ISR,s}(\alpha(Q^2),p_\gamma,y_1)}}{\dd{y} \dd{Q^2}} 
\end{align} 
or 
\begin{align} 
    \sigma_{y2}&=\dfrac{\dd[2]{\sigma^\textrm{RC(LO),ISR}(\alpha(Q^2),p_\gamma,y_2)}}{\dd{y} \dd{Q^2}}\nonumber\\
    \sigma_{y1}&=\dfrac{\dd[2]{\sigma^\textrm{RC(LO),ISR}(\alpha(Q^2),p_\gamma,y_1)}}{\dd{y} \dd{Q^2}} \,,
\end{align}
\begin{figure}
    \centering
    \includegraphics[width=\linewidth]{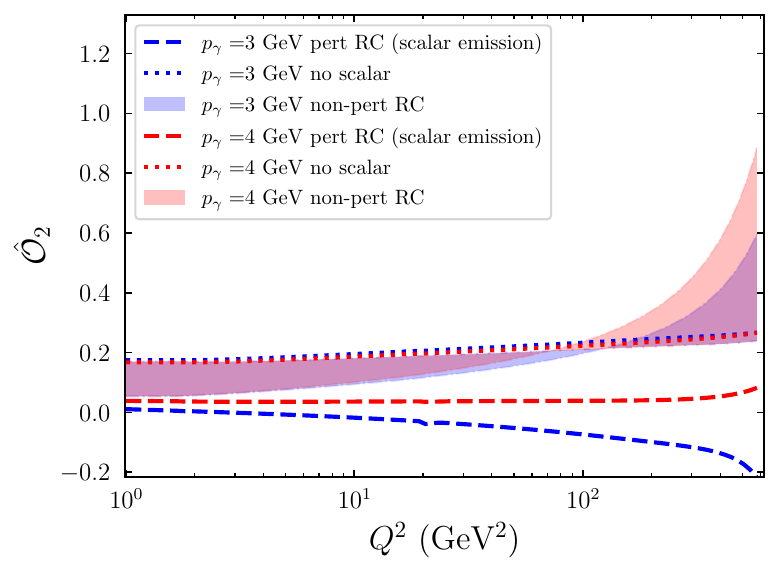}
    \caption{Observable 2 for $m_S=0.9$ GeV and $g_{se}=3\times 10^{-6}$ at specific EIC kinematics ($\sqrt{s}=140$ GeV, $E_\ell=18$ GeV, and with $y_1=0.24$ and $y_2=0.95$).}
    \label{fig:example23}
\end{figure}
or of $p_\gamma$, as in the ratio $\hat{\mathcal{O}}_3$, 
observable 3, given by 
\begin{align}\label{eq:ratio2}
    \hat{\mathcal{O}}_3&=
    \frac{\sigma_{p_{\gamma,2}}}{\sigma_{p_{\gamma,1}}}
\end{align} 
with either 
\begin{align} 
    \sigma_{p_{\gamma,2}}&=\dfrac{\dd[2]{\sigma^\textrm{RC(LO),ISR,s}(\alpha(Q^2),p_{\gamma,2},y)}}{\dd{y} \dd{Q^2}}\nonumber\\ 
    \sigma_{p_{\gamma,1}}&=\dfrac{\dd[2]{\sigma^\textrm{RC(LO),ISR,s}(\alpha(Q^2),p_{\gamma,1},y)}}{\dd{y} \dd{Q^2}}
\end{align}    
or 
\begin{align} 
    \sigma_{p_{\gamma,2}}&=\dfrac{\dd[2]{\sigma^\textrm{RC(LO),ISR}(\alpha(Q^2),p_{\gamma,2},y)}}{\dd{y} \dd{Q^2}}\nonumber\\
    \sigma_{p_{\gamma,1}}&=\dfrac{\dd[2]{\sigma^\textrm{RC(LO),ISR}(\alpha(Q^2),p_{\gamma,1},y)}}{\dd{y} \dd{Q^2}} \,.
\end{align}
These observables have been considered for the perturbative and nonperturbative models for the LDFs and LFFs separately. 
In the first case [observable 2,~\cref{eq:ratio1}], two different choices for $y$ are used, and the $p_\gamma$ range is taken to match the more restrictive kinematics of the lower $y$ value, as shown in \cref{fig:example23}. This restriction on the $p_\gamma$ range severely limits the observable effects of the scalar radiation, indicating
that this ratio is not useful 
in searching for signs of exotic particle emission.

\begin{figure}[hbt!]
    \centering
    \includegraphics[width=\linewidth]{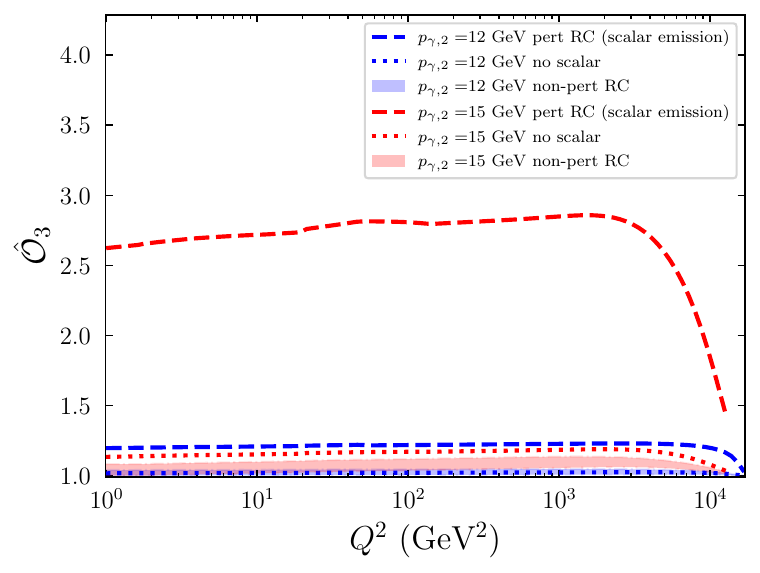}
    \caption{Observable 3 for $p_{\gamma,1}=10$ GeV, $m_S=0.9$ GeV, and $g_{se}=3\times 10^{-6}$ at specific EIC kinematics ($\sqrt{s}=140$ GeV, $E_\ell=18$ GeV, and $y=0.95$).}
    \label{fig:example33}
\end{figure}

Turning to observable 3, \cref{eq:ratio2}, 
where a fixed $y$ value 
but different $p_\gamma$ 
values 
are chosen, as 
illustrated in 
\cref{fig:example33}, 
we see the shift due to the possible scalar emission 
with a more energetic photon is highly emphasized compared to the lower energy photon emission, with a clear separation between the scalar emission model and QED-only DIS with ISR. The nonperturbative model results are also suppressed in this ratio comparison, as seen by the highlighted bands lying close to the bottom of the plot, near a ratio value of $1.$

\begin{figure}[hbt!]
    \centering
    \includegraphics[width=\linewidth]{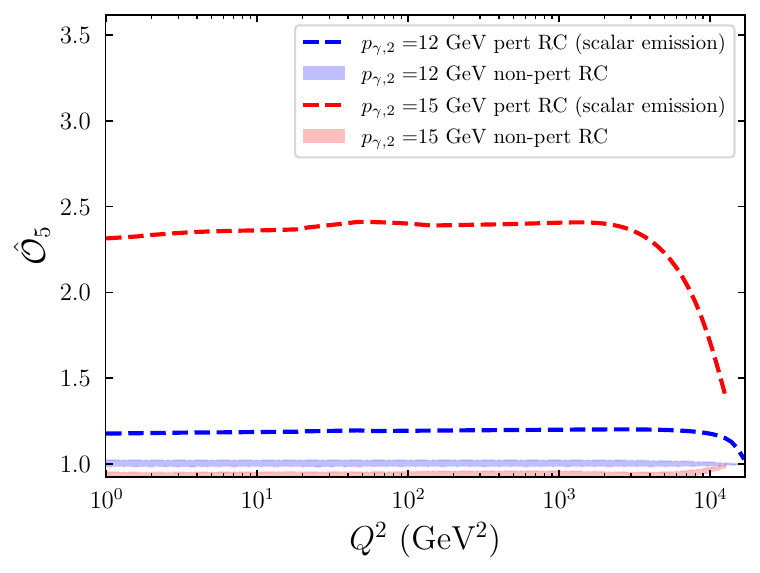}
    \caption{Observable 5 for $p_{\gamma,1}=10$ GeV, $m_S=0.9$ GeV and $g_{se}=3\times 10^{-6}$ at specific EIC kinematics ($\sqrt{s}=140$ GeV, $E_\ell=18$ GeV, and $y=0.95$).}
    \label{fig:example53}
\end{figure}

Finally, 
we consider ``super ratios'' of the 
$\hat{\cal O}_2$ and $\hat{\cal O}_3$ 
observables we have considered thus far. 
That is, we have $\hat{\mathcal{O}}_4$, observable 4, 
\begin{align}\label{eq:superratio1}
    \hat{\mathcal{O}}_4&=
    \frac{\hat{\mathcal{O}}_{2,\textrm{Scalar}}}{\hat{\mathcal{O}}_{2,\textrm{QED Only}}}=\frac{\hat{\mathcal{O}}_{1}(y_2)}{\hat{\mathcal{O}}_{1}(y_1)}
\end{align}
and $\hat{\mathcal{O}}_5$, observable 5, 
\begin{align}\label{eq:superratio2}
    \hat{\mathcal{O}}_5&=
    \frac{\hat{\mathcal{O}}_{3,\textrm{Scalar}}}{\hat{\mathcal{O}}_{3,\textrm{QED Only}}}=\frac{\hat{\mathcal{O}}_{1}(p_{\gamma,2})}{\hat{\mathcal{O}}_{1}(p_{\gamma,1})} \,. 
\end{align}
Observable 4, the super ratio in $y$, is not a useful discriminant, 
so that it is not shown, but observable 5 is a much more effective one, 
in that it 
completely minimizes the effects of the nonperturbative LDF and can help identify the scalar behavior. This is shown
in \cref{fig:example53}, where the nonperturbative effects are barely visible along the $Q^2$ axis.

\begin{figure}[hbt!]
    \centering
    \includegraphics[width=\linewidth]{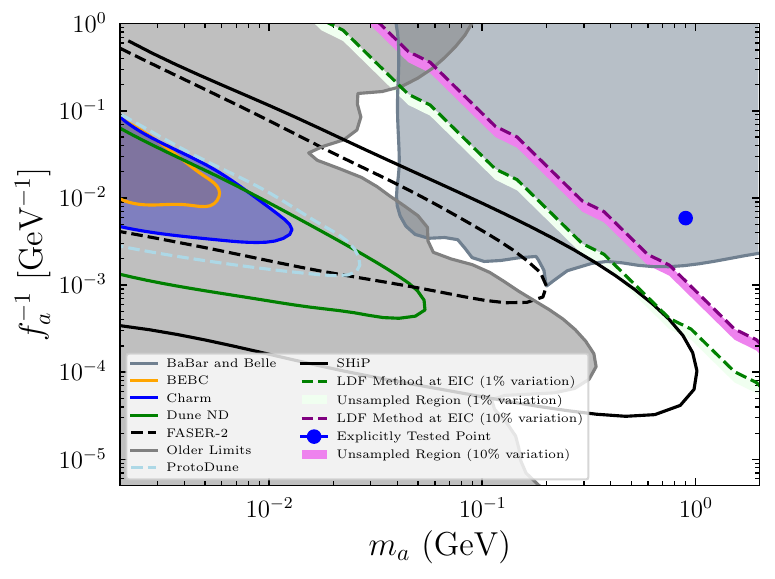}
    \caption{A facsimile of the left panel of Fig.~9 in~\cite{ema2025longlivedaxionlikeparticlestau}, 
    where we note $f_a^{-1}=g_{se}/m_e$, since our methods do not distinguish
    a scalar from an axion, along with the sensitivity 
    limits from the current work for the 
    kinematics of $y=0.95,$ $Q^2=0.75 Q^2_{\max},$ and $p_\gamma=0.9 p_{\gamma,\max}.$ Our ``LDF method''  
    results 
    show by how much 
    the computed BSM 
    cross sections can differ from the SM ones in
    the noted kinematic region, through their ratios, with 
    the associated soft bands indicating the possible variation 
    due to the limited resolution in the sampling of
    the BSM model parameter space. The single blue point indicates the mass and coupling used in the studies of 
    Figs.~\ref{fig:xsec}-\ref{fig:example53}.
    The compilation of 
    Ema \textit{et al}.~\cite{ema2025longlivedaxionlikeparticlestau} 
    shows excluded regions on the parameter space from existing experiments 
    as shaded regions (typically at 90\% CL), and the solid and dashed
    lines are forecasts. In all cases, these collected results
    and forecasts limit a produced scalar $\phi$ through nonobservation of 
    $\phi\to e^+ e^-$. The sensitivity limits we show also constrain models in which $\phi$ could decay to other dark-sector particles, rendering them distinct from 
    all the others shown.     
    We refer to the text for all
    details.}
    \label{fig:emacomp}
\end{figure}

Overall, since our limits apply to axionlike
particles as well, we can compare to the 
analysis of limits on MeV-GeV scale 
axions 
from Ema \textit{et al}.~\cite{ema2025longlivedaxionlikeparticlestau}. 
Those authors consider limits on 
axion-lepton coupling in various flavorful
scenarios. In that case it turns out that 
the couplings in the $\mu$-phobic axion
scenario, with $g_{a\mu} =0$ and $g_{a e}=g_{a\tau} =m_e /f_a$, 
where $f_a$ is a parameter to be determined, 
are the most poorly constrained. This is shown
in the left panel of their Fig.~9~\cite{ema2025longlivedaxionlikeparticlestau}, which is 
roughly 
reproduced for a proof-of-principle comparison in~\cref{fig:emacomp}. Our examples in 
Figs.~\ref{fig:xsec}-\ref{fig:example53} employ
$g_{se}=3\times 10^{-6}$ or $f_{a}^{-1}=6\times 10^{-3}$ GeV$^{-1}$ and a scalar mass of 
$0.9\,\rm GeV$, which is indicated as a single blue point. If
we set aside the possibility of dark decays of the scalar, this lies above the border of the parameter space excluded by the \textit{BABAR}~\cite{BaBar:2020jma} and Belle~\cite{Belle:2022gbl} data. Even so, these examples are illustrative of the effect a scalar candidate would have on the cross section, and sample data point calculations for $g_{se}\approx 3\times 10^{-7},$ or $f_{a}^{-1}\approx 6\times 10^{-4},$ show the trends would be similar, but slightly smaller in magnitude. This choice for $g_{se}$ around $10^{-7}$ falls outside the excluded region while remaining within the sensitivity limits from this study. The reasoning for the original parameter choice is explained in~\cref{sec:kinchoice}.

Here we work in the kinematics of $y=0.95,$ $Q^2=0.75 Q^2_{\max},$ and $p_\gamma=0.9 p_{\gamma,\max}$ to focus on a region with the largest effects. 
For this case, the limits from this study, a general spin 0 leptophilic candidate, are added in dashed green and purple. 
These represent changes at the level of $1\%$ or $10\%$ in the cross section
ratio $\hat{\mathcal{O}}_1$, respectively, meaning that the cross section with scalar emission changes by at least that amount from the SM-only cross section.
In contrast to the limits and forecasts in Ema \textit{et al}.~\cite{ema2025longlivedaxionlikeparticlestau}, 
the dashed lines from this study represent a 
deviation from a  cross section ratio of unity 
across a smooth region. The variations chosen for this plot 
are simply illustrative of the parameter space this method can probe, and any expected deviation from the SM-only prediction can be used to point to a discovery. 
The lighter bands represent the range of possible testable candidates which could fall in the given variation level, as the given limits were found by dividing the mass and coupling phase space into $50$ points each. The cross section ratio 
was calculated at each pair for the given kinematics, and the deviation was checked to be at least the desired size. Thus, there would be 
some mass and coupling points that fall between the chosen ones, and the 
band thickness follow from determining where 
the 
tested parameter choice was found to 
have insufficient deviation from the SM-only prediction. 
When considering the expected integrated luminosity of the EIC to be $\approx 10$ fb$^{-1}$ per year~\cite{Abdul_Khalek_2022_short}, the number of events expected can be found to be roughly $10^9,$  
provides a comfortable margin for detector limitations, i.e., 
imperfect detector efficiency, to ensure strong statistics.

\section{Summary and Outlook\label{sec:sum}}
In this paper we have set out a new search 
strategy at accelerators 
for the emission of non-SM particles
of varying mass and spin through the 
detection of anomalous energy loss, i.e., 
through the determination of the modification 
of the cross section that their emission 
would generate. This new strategy is 
made possible through the use of the 
joint QED and QCD factorization of deeply
inelastic, electron-proton scattering~\cite{liu:2020rvc,liu2021new,cammarota2025factorizedqedqcdcontribution}, because 
it provides a systematic treatment of 
the radiative corrections that appear. 
This approach eliminates the 
parameter-dependent corrections that 
can vary from framework to framework
in earlier work, at the price of introducing 
universal, nonperturbative LDF and LFF functions to 
be determined from DIS experiments. Here we have
leveraged the sensitivity of the resulting cross section 
to new, light, weakly coupled particle emission through the additional 
requirement of 
detecting an energetic photon from ISR. 
In this paper we have studied the particular case of 
spin $0$ particle emission, because our approach, at its
current level of refinement, does not determine
the parity of that emitted, energetic particle, to show
that in the case of an EIC study we can expect to probe previously unconstrained BSM parameter space, particularly at 
MeV-GeV scales, concretely. 

We expect the forthcoming EIC to give us new 
insight into the physics of the femtometer scale. 
At this new frontier 
we can probe not only 
the intricate color, spin, and 
flavor correlations of the strongly interacting quarks and
gluons within nucleons~\cite{Accardi:2012qut}, but we also 
see that it opens the possibility of studying non-SM 
physics 
in the challenging MeV-GeV regime of new particle masses,  
for which other sorts of 
experiments are under development
worldwide~\cite{Berlin:2018bsc,NA64:2025ddk,LDMX:2025bog,Yang:2025yvl}. 

The possibility of 
displaced vertex searches have already been discussed 
for study at the EIC~\cite{Davoudiasl:2023pkq,Balkin:2023gya}, and we would like to emphasize that the anomalous energy loss search we advocate
allows us to probe a broader class of models. The displaced vertex search for a new scalar, say, has 
a sensitivity that scales as $g_{se}^4$, and it requires that the 
new particle materializes into a dilepton pair. However, in 
an analogous anomalous energy loss search (here our LDF method), which scales as $g_{se}^2$, the new particle can decay to other, unspecified, particles of the hidden sector. When the coupling to other dark-sector particles is larger than the SM coupling, these other decay modes are more likely, making it harder to detect the decay of the new particle into a dilepton pair in a displaced vertex search. In this manner the anomalous energy 
loss method probes a much broader class of models and opens access
to weaker $g_{se}$ couplings as well.

Although we have focused on probing the possibility of 
spin $0$ particle emission, many sorts of particle probes
and tests are possible. We have noted these 
broader prospects throughout the paper, 
and we have 
enumerated the quantum numbers and certain interactions 
of the dark matter particles
accessible to us in Table~\ref{tab:dmlagrangians}.  
Here, in summary, we collect some of the possibilities in an organized way. 
We believe the MeV-GeV mass scale opens a number of 
interesting questions 
to explore, even if 
the various BSM particle candidates could 
appear with an enormous range of masses. 
Particularly, 
\begin{enumerate}[(i)]
\item We have discussed how the comparison of dimension 4 and 
5 spin 0 particle-electron interactions are sensitive to the parity 
of the spin 0 particle. In particular, in the axionlike case, the 
possibility of the process $\gamma^* \to a \gamma$, radiated from an 
electron line, would seem to be required to connect the two 
descriptions precisely. This process, in turn, feeds into 
the nonperturbative LDF, thus yielding a BSM contribution to 
its construction, and showing that this object can be 
different in the SM and BSM cases. 

\item We observe that our current RC(LO) analysis would 
only allow us to fit 
for nonperturbative SM and BSM LDFs separately through 
differences in their respective splitting functions. However, 
if we were to extend our analysis to NLO in the hard part, 
the BSM particle would appear explicitly and thus provide
additional distinguishing features for that analysis. 

\item Although we have focused on other possibilities in this
paper, it should also be possible to study massive spin 2 
particles with our methods as well, albeit they are typically considered in 
mass regions apropos to ultralight dark matter, i.e., at sub-eV scales.

\item The joint QED and QCD factorization framework allows us 
to step beyond the cross section to probe the interactions of the 
partons in the struck hadron in an DIS event. 
Particularly, we think that the study of the energy-energy correlators 
in this context could prove a powerful way of 
probing the possibility of energy loss from hidden-sector interactions 
with gluons. 
\end{enumerate}
\medskip

We look forward to future exploration of these issues and more
at the EIC and elsewhere.
\medskip

\acknowledgments 
We gratefully acknowledge discussions with Jianwei Qiu, particularly of 
his work 
on isolated photon emission in DIS~\cite{BergerQiuPrompt}, that we utilize
here, and we also thank the 
U.S. Department of Energy, Office of Nuclear Physics, 
under Grant No. DE-FG02-96ER40989 for partial support. We would like to acknowledge 
Hooman Davoudiasl for a helpful comment.
We thank the BNL EIC Theory institute for their hospitality while this paper is in preparation.
This work has been partially funded by the Deutsche Forschungsgemeinschaft (DFG, German Research Foundation) No. 491245950.

\section*{Data Availability}
There are no publicly available research data or software supporting this manuscript. Requests for further information or data should be sent to the authors.

\appendix
\section{DETAILED KINEMATIC CHOICES}\label{sec:kinchoice}
The exact kinematic inputs were found by setting the bounds for the range of each value and dividing it into even segments. The results above show the rounded, ``clean'' numbers for reference, however, here the exact values will be listed. For the range of $y,$ it was taken as the expected EIC values, $0.01$ to $0.95,$ then divided into five segments (including the end points), so for the ratio between $y_1$ and $y_2,$ the $y_1=0.245$ exactly (all the exact $y$ values are shown in \cref{fig:pyvy}). The mass of the scalar was evenly distributed over a logarithmic space (also with 5 segments), from $10^{-5}$ GeV to the max allowed $m_S$ for a given kinematic ($E_\ell-p_\gamma$), meaning the larger $y$ limited the $p_\gamma$ and therefore the $m_S,$ so for $y=0.95,$ the largest $m_S$ possible was $9.0005\times 10^{-1}$ GeV, which was reported above at the tenth of a GeV level. Similarly, the $g_{se}$ range was evenly distributed over five segments of a logarithmic space from $10^{-9}$ to $10^{-2},$ which made the exact value for coupling above to be $g_{se}=3.162\times 10^{-6}.$ The $p_\gamma$ range was divided into five segments also, but linearly over the range from the imposed minimum of $0.6 p_{\gamma,\max}$ to $0.9 p_{\gamma,\max}.$ The minimum cutoff is discussed above, but the maximum cutoff was chosen to avoid end point issues as when $p_{\gamma}=p_{\gamma,\max},$ $Q^2_{\max}=1,$ severely limiting the viability of factorization at this scale [see \cref{eq:pqlimits}]. The exact values for each the $p_\gamma$ above are then (all in units of GeV): $15\rightarrow 15.39,$ $12\rightarrow 11.54,$ $10\rightarrow 10.26,$ $4\rightarrow 3.97,$ and $3\rightarrow 2.65,$ where we left the values at the two decimal places for simplicity. These more precise values are included here to ensure reproducibility of our results with the same kinematic choices.

\bibliography{bib,EICplusDIS}

@article{Han:2021kes,
    author = "Han, Tao and Ma, Yang and Xie, Keping",
    title = "{Quark and gluon contents of a lepton at high energies}",
    eprint = "2103.09844",
    archivePrefix = "arXiv",
    primaryClass = "hep-ph",
    reportNumber = "PITT-PACC-2104",
    doi = "10.1007/JHEP02(2022)154",
    journal = "JHEP",
    volume={2022},
    ISSN={1029-8479},
    url={http://dx.doi.org/10.1007/JHEP02(2022)154},
    number={2},
    pages = "154",
    year = "2022"
}

@article{Adler:1969gk,
    author = "Adler, Stephen L.",
    title = "{Axial vector vertex in spinor electrodynamics}",
    doi = "10.1103/PhysRev.177.2426",
    journal = "Phys. Rev.",
    volume = "177",
    pages = "2426--2438",
    year = "1969"
}

@article{Bell:1969ts,
    author = "Bell, J. S. and Jackiw, R.",
    title = "{A PCAC puzzle: $\pi^0 \to \gamma \gamma$ in the $\sigma$ model}",
    doi = "10.1007/BF02823296",
    journal = "Nuovo Cim. A",
    volume = "60",
    pages = "47--61",
    year = "1969"
}

@article{Adler:1969er,
    author = "Adler, Stephen L. and Bardeen, William A.",
    title = "{Absence of higher order corrections in the anomalous axial vector divergence equation}",
    doi = "10.1103/PhysRev.182.1517",
    journal = "Phys. Rev.",
    volume = "182",
    pages = "1517--1536",
    year = "1969"
}

@article{Adshead:2021ezw,
    author = "Adshead, Peter and Lozanov, Kaloian D.",
    title = "{Axion anomalies}",
    eprint = "2112.07645",
    archivePrefix = "arXiv",
    primaryClass = "hep-th",
    doi = "10.1007/JHEP08(2022)077",
    journal = "JHEP",
    volume={2022},
    ISSN={1029-8479},
    url={http://dx.doi.org/10.1007/JHEP08(2022)077},
    number={8},
    pages = "077",
    year = "2022"
}

@article{Peccei:1977hh,
    author = "Peccei, R. D. and Quinn, Helen R.",
    title = "{CP Conservation in the Presence of Instantons}",
    reportNumber = "ITP-568-STANFORD",
    doi = "10.1103/PhysRevLett.38.1440",
    journal = "Phys. Rev. Lett.",
    volume = "38",
    pages = "1440--1443",
    year = "1977"
}

@article{Berlin:2023ubt,
    author = "Berlin, Asher and Millar, Alexander J. and Trickle, Tanner and Zhou, Kevin",
    title = "{Physical signatures of fermion-coupled axion dark matter}",
    eprint = "2312.11601",
    archivePrefix = "arXiv",
    primaryClass = "hep-ph",
    reportNumber = "FERMILAB-PUB-23-779-SQMS-T, SLAC-PUB-17758",
    doi = "10.1007/JHEP05(2024)314",
    volume={2024},
    ISSN={1029-8479},
    url={http://dx.doi.org/10.1007/JHEP05(2024)314},
    number={5},
    journal = "JHEP",
    pages = "314",
    year = "2024"
}

@article{Kogut:1969xa,
    author = "Kogut, John B. and Soper, Davison E.",
    title = "{Quantum Electrodynamics in the Infinite Momentum Frame}",
    reportNumber = "SLAC-PUB-0680",
    doi = "10.1103/PhysRevD.1.2901",
    journal = "Phys. Rev. D",
    volume = "1",
    pages = "2901--2913",
    year = "1970"
}

@article{deFlorian:2015ujt,
    author = "de Florian, Daniel and Sborlini, Germ{\'a}n F. R. and Rodrigo, Germ{\'a}n",
    title = "{QED corrections to the Altarelli{\textendash}Parisi splitting functions}",
    eprint = "1512.00612",
    archivePrefix = "arXiv",
    primaryClass = "hep-ph",
    reportNumber = "ICAS-03-15, IFIC-15-81",
    doi = "10.1140/epjc/s10052-016-4131-8",
    journal = "Eur. Phys. J. C",
    volume = "76",
    number = "5",
    pages = "282",
    year = "2016"
}

@article{deFlorian:2016gvk,
    author = "de Florian, Daniel and Sborlini, Germ{\'a}n F. R. and Rodrigo, Germ{\'a}n",
    title = "{Two-loop QED corrections to the Altarelli-Parisi splitting functions}",
    eprint = "1606.02887",
    archivePrefix = "arXiv",
    ISSN={1029-8479},
    primaryClass = "hep-ph",
    reportNumber = "ICAS-09-16, IFIC-15-88",
    doi = "10.1007/JHEP10(2016)056",
    url={http://dx.doi.org/10.1007/JHEP10(2016)056},
    journal = "JHEP",
    volume = "2016",
    number="10",
    pages = "056",
    year = "2016"
}

@BOOK{Maggiore2007gwte.book.....M,
       author = {{Maggiore}, Michele},
        title = "{Gravitational Waves: Volume 1: Theory and Experiments}",
         year = 2007,
          doi = {10.1093/acprof:oso/9780198570745.001.0001},
      publisher="Oxford University Press",          
       adsurl = {https://ui.adsabs.harvard.edu/abs/2007gwte.book.....M}
}

@misc{Electron-PositronAlliance:2025fhk,
    author = "Bossi, Hannah and others",
    collaboration = "Electron-Positron Alliance",
    title = "{Energy Correlators from Partons to Hadrons: Unveiling the Dynamics of the Strong Interactions with Archival ALEPH Data}",
    eprint = "2511.00149",
    archivePrefix = "arXiv",
    primaryClass = "hep-ph",
    reportNumber = "MITP-25-057, MITHIG-MOD-24-001",
    month = "10",
    year = "2025"
}

@article{Fierz:1939ix,
    author = "Fierz, M. and Pauli, W.",
    title = "{On relativistic wave equations for particles of arbitrary spin in an electromagnetic field}",
    doi = "10.1098/rspa.1939.0140",
    journal = "Proc. Roy. Soc. Lond. A",
    volume = "173",
    pages = "211--232",
    year = "1939"
}

@article{Lin:2025hka,
    author = "Lin, Huey-Wen",
    title = "{Mapping parton distributions of hadrons with lattice QCD}",
    eprint = "2506.05025",
    archivePrefix = "arXiv",
    primaryClass = "hep-lat",
    reportNumber = "MSUHEP-24-012",
    doi = "10.1016/j.ppnp.2025.104177",
    journal = "Prog. Part. Nucl. Phys.",
    volume = "144",
    pages = "104177",
    year = "2025"
}

@inproceedings{Chou:2022luk,
    author = "Chou, Aaron S. and others",
    title = "{Snowmass Cosmic Frontier Report}",
    booktitle = "{Snowmass 2021}",
    eprint = "2211.09978",
    archivePrefix = "arXiv",
    primaryClass = "hep-ex",
    reportNumber = "UCI-HEP-TR-2022-26, FERMILAB-FN-1217-PPD-SCD",
    doi = "10.2172/1908206",
    month = "11",
    year = "2022"
}

@inproceedings{Alarcon:2022ero,
    author = "Alarcon, Ricardo and others",
    title = "{Electric dipole moments and the search for new physics}",
    booktitle = "{Snowmass 2021}",
    eprint = "2203.08103",
    archivePrefix = "arXiv",
    primaryClass = "hep-ph",
    doi="10.48550/arXiv.2203.08103",
    month = "3",
    year = "2022"
}

@article{Tsai:1961zz,
    author = "Tsai, Yung-Su",
    title = "{Radiative Corrections to Electron-Proton Scattering}",
    doi = "10.1103/PhysRev.122.1898",
    journal = "Phys. Rev.",
    volume = "122",
    pages = "1898--1907",
    year = "1961"
}

@inproceedings{Tsai:1971qi,
    author = "Tsai, Yung-Su",
    title = "{RADIATIVE CORRECTIONS TO ELECTRON SCATTERINGS}",
    reportNumber = "SLAC-PUB-0848, SLAC-PUB-848",
    month = "1",
    year = "1971"
}

@article{Bardin:1988by,
    author = "Bardin, D. Yu. and Burdik, C. and Khristova, P. Ch. and Riemann, T.",
    title = "{ELECTROWEAK RADIATIVE CORRECTIONS TO DEEP INELASTIC SCATTERING AT HERA. NEUTRAL CURRENT SCATTERING}",
    reportNumber = "PHE-88-15",
    doi = "10.1007/BF01557676",
    journal = "Z. Phys. C",
    volume = "42",
    pages = "679",
    year = "1989"
}

@article{deRham:2014zqa,
    author = "de Rham, Claudia",
    title = "{Massive Gravity}",
    eprint = "1401.4173",
    archivePrefix = "arXiv",
    primaryClass = "hep-th",
    doi = "10.12942/lrr-2014-7",
    journal = "Living Rev. Rel.",
    volume = "17",
    pages = "7",
    year = "2014"
}

@ARTICLE{LIGO_2016PhRvX...6d1015A,
       author = "{Abbott}, B.~P. and others",
        title = "{Binary Black Hole Mergers in the First Advanced LIGO Observing Run}",
      journal = {Physical Review X},
         year = 2016,
        month = oct,
       volume = {6},
       number = {4},
          eid = {041015},
        pages = {041015},
          doi = {10.1103/PhysRevX.6.041015},
archivePrefix = {arXiv},
       eprint = {1606.04856},
 primaryClass = {gr-qc},
       adsurl = {https://ui.adsabs.harvard.edu/abs/2016PhRvX...6d1015A}
}

@article{Davoudiasl:2023pkq,
    author = "Davoudiasl, Hooman and Marcarelli, Roman and Neil, Ethan T.",
    title = "{Displaced signals of hidden vectors at the Electron-Ion Collider}",
    eprint = "2307.00102",
    archivePrefix = "arXiv",
    primaryClass = "hep-ph",
    doi = "10.1103/PhysRevD.108.075017",
    journal = "Phys. Rev. D",
    volume = "108",
    number = "7",
    pages = "075017",
    year = "2023"
}

@article{Balkin:2023gya,
    author = "Balkin, Reuven and Hen, Or and Li, Wenliang and Liu, Hongkai and Ma, Teng and Soreq, Yotam and Williams, Mike",
    title = "{Probing axion-like particles at the Electron-Ion Collider}",
    eprint = "2310.08827",
    archivePrefix = "arXiv",
    primaryClass = "hep-ph",
    doi = "10.1007/JHEP02(2024)123",
    journal = "JHEP",
    volume = "2024",
    number = "123",
    year = "2024"
}

@article{BaBar:2020jma,
    author = "Lees, J. P. and others",
    collaboration = "BABAR Collaboration",
    title = "{Search for a Dark Leptophilic Scalar in $e^+e^-$ Collisions}",
    eprint = "2005.01885",
    archivePrefix = "arXiv",
    primaryClass = "hep-ex",
    reportNumber = "BABAR-PUB-20/003, SLAC-PUB-17526",
    doi = "10.1103/PhysRevLett.125.181801",
    journal = "Phys. Rev. Lett.",
    volume = "125",
    number = "18",
    pages = "181801",
    year = "2020"
}

@article{Belle:2022gbl,
    author = "Biswas, D. and others",
    collaboration = "Belle Collaboration",
    title = "{Search for a dark leptophilic scalar produced in association with {\ensuremath{\tau}}+{\ensuremath{\tau}}- pair in e+e- annihilation at center-of-mass energies near 10.58~GeV}",
    eprint = "2207.07476",
    archivePrefix = "arXiv",
    primaryClass = "hep-ex",
    reportNumber = "BELLE-CONF-2201, Belle Preprint 2023-13, KEK Preprint 2023-17",
    doi = "10.1103/PhysRevD.109.032002",
    journal = "Phys. Rev. D",
    volume = "109",
    number = "3",
    pages = "032002",
    year = "2024"
}

@article{Blas:2024kps,
    author = "Blas, Diego and Carlton, John and McCabe, Christopher",
    title = "{Massive graviton dark matter searches with long-baseline atom interferometers}",
    eprint = "2412.14282",
    archivePrefix = "arXiv",
    primaryClass = "hep-ph",
    reportNumber = "KCL-PH-TH/2024-69, AION-REPORT/2024-08",
    doi = "10.1103/zxtk-bwnf",
    journal = "Phys. Rev. D",
    volume = "111",
    number = "11",
    pages = "115020",
    year = "2025"
}

@misc{Moult:2025nhu,
    author = "Moult, Ian and Zhu, Hua Xing",
    title = "{Energy Correlators: A Journey From Theory to Experiment}",
    eprint = "2506.09119",
    archivePrefix = "arXiv",
    primaryClass = "hep-ph",
    month = "6",
    year = "2025"
}

@article{Accardi:2012qut,
    author = "Accardi, A. and others",
    editor = "Deshpande, A. and Meziani, Z. E. and Qiu, J. W.",
    title = "{Electron Ion Collider: The Next QCD Frontier}: {Understanding the glue that binds us all}",
    eprint = "1212.1701",
    archivePrefix = "arXiv",
    primaryClass = "nucl-ex",
    reportNumber = "BNL-98815-2012-JA, JLAB-PHY-12-1652",
    doi = "10.1140/epja/i2016-16268-9",
    journal = "Eur. Phys. J. A",
    volume = "52",
    number = "9",
    pages = "268",
    year = "2016"
}

@article{Berlin:2018bsc,
    author = "Berlin, Asher and Blinov, Nikita and Krnjaic, Gordan and Schuster, Philip and Toro, Natalia",
    title = "{Dark Matter, Millicharges, Axion and Scalar Particles, Gauge Bosons, and Other New Physics with LDMX}",
    eprint = "1807.01730",
    archivePrefix = "arXiv",
    primaryClass = "hep-ph",
    reportNumber = "FERMILAB-PUB-18-310-A, SLAC-PUB-17297",
    doi = "10.1103/PhysRevD.99.075001",
    journal = "Phys. Rev. D",
    volume = "99",
    number = "7",
    pages = "075001",
    year = "2019"
}

@misc{NA64:2025ddk,
    author = "Andreev, Yu. M. and others",
    collaboration = "NA64 Collaboration",
    title = "{Searching for Light Dark Matter and Dark Sectors with the NA64 experiment at the CERN SPS}",
    eprint = "2505.14291",
    archivePrefix = "arXiv",
    primaryClass = "hep-ex",
    month = "5",
    year = "2025"
}

@misc{LDMX:2025bog,
      title="{LDMX -- The Light Dark Matter eXperiment}", 
      author={Torsten Akesson and others},
      year={2025},
      eprint={2508.11833},
      archivePrefix={arXiv},
      primaryClass={hep-ex},     
}

@inproceedings{Yang:2025yvl,
    author = "Yang, Haijun",
    title = "{DarkSHINE: Search for Light Dark Matter at the SHINE Facility in Shanghai}",
    booktitle = "{32nd International Symposium on Lepton Photon Interactions at High Energies}: {Lepton-Photon 2025}",
    eprint = "2512.13301",
    archivePrefix = "arXiv",
    primaryClass = "hep-ex",
    month = "12",
    year = "2025"
}

@article{Gribov:1972ri,
    author = "Gribov, V. N. and Lipatov, L. N.",
    title = "{Deep inelastic e p scattering in perturbation theory}",
    reportNumber = "IPTI-381-71",
    journal = "Sov. J. Nucl. Phys.",
    volume = "15",
    pages = "438--450",
    year = "1972",
    url={https://www.osti.gov/biblio/4621331}
}

@article{Altarelli:1977zs,
    author = "Altarelli, Guido and Parisi, G.",
    title = "{Asymptotic Freedom in Parton Language}",
    reportNumber = "LPTENS-77-6",
    doi = "10.1016/0550-3213(77)90384-4",
    journal = "Nucl. Phys. B",
    volume = "126",
    pages = "298--318",
    year = "1977"
}

@book{Collins:2011zzd,
    author = "Collins, John",
    title = "{Foundations of Perturbative QCD}",
    doi = "10.1017/9781009401845",
    isbn = "978-1-009-40184-5, 978-1-009-40183-8, 978-1-009-40182-1",
    publisher = "Cambridge University Press",
    volume = "32",
    year = "2011"
}

@article{liu:2020rvc,
    author = "Liu, Tianbo and Melnitchouk, W. and Qiu, Jian-Wei and Sato, N.",
    title = "{Factorized approach to radiative corrections for inelastic lepton-hadron collisions}",
    eprint = "2008.02895",
    archivePrefix = "arXiv",
    primaryClass = "hep-ph",
    reportNumber = "JLAB-THY-20-3233",
    doi = "10.1103/PhysRevD.104.094033",
    journal = "Phys. Rev. D",
    volume = "104",
    number = "9",
    pages = "094033",
    year = "2021"
}

@article{liu2021new,
	doi = {10.1007/jhep11(2021)157},
  
	url = {https://doi.org/10.1007%2Fjhep11%282021%29157},
  
	year = 2021,
	month = {nov},
  
	publisher = {Springer Science and Business Media {LLC}
},
  
	volume = {2021},
  
	number = {11},
  
	author = {Tianbo Liu and W. Melnitchouk and Jian-Wei Qiu and N. Sato},
  
	title = {A new approach to semi-inclusive deep-inelastic scattering with {QED} and {QCD} factorization},
  
	journal = {Journal of High Energy Physics}
}

@ARTICLE{BergerQiuPrompt,
       author = {{Berger}, Edmond L. and {Qiu}, Jianwei},
        title = "{Understanding the cross section for isolated prompt photon production}",
      journal = {Physics Letters B},
         year = 1990,
        month = oct,
       volume = {248},
       number = {3-4},
        pages = {371-377},
          doi = {10.1016/0370-2693(90)90308-S},
}

@article{cammarota2025factorizedqedqcdcontribution,
  title = {Factorized QED and QCD contribution to deeply inelastic scattering},
  author = {Cammarota, Justin and Qiu, Jian-Wei and Watanabe, Kazuhiro and Zhang, Jia-Yue},
  journal = {Phys. Rev. D},
  volume = {112},
  issue = {5},
  pages = {056007},
  numpages = {20},
  year = {2025},
  month = {Sep},
  publisher = {American Physical Society},
  doi = {10.1103/2d8y-ljwx},
  url = {https://link.aps.org/doi/10.1103/2d8y-ljwx}
}

@article{Abdul_Khalek_2022_short,
   title={Science Requirements and Detector Concepts for the Electron-Ion Collider},
   volume={1026},
   ISSN={0375-9474},
   url={http://dx.doi.org/10.1016/j.nuclphysa.2022.122447},
   DOI={10.1016/j.nuclphysa.2022.122447},
   journal={Nuclear Physics A},
   publisher={Elsevier BV},
   author={Abdul Khalek, R. and others},
   year={2022},
   month=oct, pages={122447} }

@article{Anastasi_2017_short,
   title="{Measurement of the running of the fine structure constant below 1 GeV with the KLOE detector}",
   volume={767},
   ISSN={0370-2693},
   url={http://dx.doi.org/10.1016/j.physletb.2016.12.016},
   DOI={10.1016/j.physletb.2016.12.016},
   journal={Physics Letters B},
   publisher={Elsevier BV},
   author={Anastasi, A. and others},
   year={2017},
   month=apr, pages={485–492} }

@article{Collins:1985ue,
    author = "Collins, John C. and Soper, Davison E. and Sterman, George F.",
    title = "{Factorization for Short Distance Hadron - Hadron Scattering}",
    reportNumber = "OITS-287",
    doi = "10.1016/0550-3213(85)90565-6",
    journal = "Nucl. Phys. B",
    volume = "261",
    pages = "104--142",
    year = "1985"
}

@article{COLLINS1982445,
title = {Parton distribution and decay functions},
journal = {Nuclear Physics B},
volume = {194},
number = {3},
pages = {445-492},
year = {1982},
issn = {0550-3213},
doi = {https://doi.org/10.1016/0550-3213(82)90021-9},
url = {https://www.sciencedirect.com/science/article/pii/0550321382900219},
author = {John C. Collins and Davison E. Soper}
}

@article{ema2025longlivedaxionlikeparticlestau,
    author = "Ema, Yohei and Fox, Patrick J. and Hostert, Matheus and Menzo, Tony and Pospelov, Maxim and Ray, Anupam and Zupan, Jure",
    title = "{Long-lived axionlike particles from tau decays}",
    eprint = "2507.15271",
    archivePrefix = "arXiv",
    primaryClass = "hep-ph",
    reportNumber = "CERN-TH-2025-123, FERMILAB-PUB-25-0408-T, N3AS-25-011",
    doi = "10.1103/51gx-m32t",
    journal = "Phys. Rev. D",
    volume = "112",
    number = "11",
    pages = "115028",
    year = "2025"
}

@article{Turner:1989vc,
    author = "Turner, Michael S.",
    title = "{Windows on the Axion}",
    reportNumber = "FERMILAB-CONF-89-104-A",
    doi = "10.1016/0370-1573(90)90172-X",
    journal = "Phys. Rept.",
    volume = "197",
    pages = "67--97",
    year = "1990"
}

@misc{donoghue2017epfllecturesgeneralrelativity,
      title="{EPFL Lectures on General Relativity as a Quantum Field Theory}", 
      author={John F. Donoghue and Mikhail M. Ivanov and Andrey Shkerin},
      year={2017},
      eprint={1702.00319},
      archivePrefix={arXiv},
      primaryClass={hep-th},       
}

@article{RevModPhys.41.205,
  title = {Radiative Corrections to Elastic and Inelastic $\mathrm{ep}$ and $\mathrm{up}$ Scattering},
  author = {Mo, L. W. and Tsai, Y. S.},
  journal = {Rev. Mod. Phys.},
  volume = {41},
  issue = {1},
  pages = {205--235},
  numpages = {0},
  year = {1969},
  month = {Jan},
  publisher = {American Physical Society},
  doi = {10.1103/RevModPhys.41.205},
  url = {https://link.aps.org/doi/10.1103/RevModPhys.41.205}
}

@article{Bardin:1976qa,
    author = "Bardin, D. Yu. and Shumeiko, N. M.",
    title = "{An Exact Calculation of the Lowest Order Electromagnetic Correction to the Elastic Scattering}",
    reportNumber = "JINR-P2-10113",
    doi = "10.1016/0550-3213(77)90213-9",
    journal = "Nucl. Phys.",
    volume = "B127",
    pages = "242--258",
    year = "1977"
}

@article{Bardin:1989vz,
    author = "Bardin, D. Yu. and Burdik, K. C. and Khristova, P. Kh. and Riemann, T.",
    title = "{Electroweak radiative corrections to deep inelastic scattering at HERA. Charged current scattering}",
    reportNumber = "JINR-E2-89-145",
    doi = "10.1007/BF01548593",
    journal = "Z. Phys. C",
    volume = "44",
    pages = "149",
    year = "1989"
}

@article{Kripfganz:1990vm,
    author = "Kripfganz, J. and Mohring, H. J. and Spiesberger, H.",
    title = "{Higher order leading logarithmic QED corrections to deep inelastic e p scattering at very high-energies}",
    reportNumber = "DESY-90-096",
    doi = "10.1007/BF01549704",
    journal = "Z. Phys. C",
    volume = "49",
    pages = "501--510",
    year = "1991"
}

@article{Spiesberger:1994dm,
    author = "Spiesberger, H.",
    title = "{QED radiative corrections for parton distributions}",
    eprint = "hep-ph/9412286",
    archivePrefix = "arXiv",
    reportNumber = "BI-TP-94-60, BI-TP.94-60",
    doi = "10.1103/PhysRevD.52.4936",
    journal = "Phys. Rev. D",
    volume = "52",
    pages = "4936--4940",
    year = "1995"
}

@article{Blumlein:2007kx,
    author = "Blumlein, Johannes and Kawamura, Hiroyuki",
    title = "{Universal higher order singlet QED corrections to unpolarized lepton scattering}",
    eprint = "hep-ph/0701019",
    archivePrefix = "arXiv",
    reportNumber = "DESY-04-064, SFB-CPP-06-63",
    doi = "10.1140/epjc/s10052-007-0300-0",
    journal = "Eur. Phys. J. C",
    volume = "51",
    pages = "317--333",
    year = "2007"
}

@article{Badelek:1994uq,
    author = "Badelek, B. and Bardin, Dmitri Yu. and Kurek, K. and Scholz, C.",
    title = "{Radiative correction schemes in deep inelastic muon scattering}",
    eprint = "hep-ph/9403238",
    archivePrefix = "arXiv",
    reportNumber = "TSL-ISV-94-0092",
    doi = "10.1007/BF01579633",
    journal = "Z. Phys. C",
    volume = "66",
    pages = "591--600",
    year = "1995"
}

@article{PhysRevD.73.036007,
  title = {Discrete symmetries on the light front and a general relation connecting the nucleon electric dipole and anomalous magnetic moments},
  author = {Brodsky, Stanley J. and Gardner, Susan and Hwang, Dae Sung},
  journal = {Phys. Rev. D},
  volume = {73},
  issue = {3},
  pages = {036007},
  numpages = {13},
  year = {2006},
  month = {Feb},
  publisher = {American Physical Society},
  doi = {10.1103/PhysRevD.73.036007},
  url = {https://link.aps.org/doi/10.1103/PhysRevD.73.036007}
}

@article{Guo:2025qnz,
    author = "Guo, Yuxun and Vogelsang, Werner and Yuan, Feng and Zhao, Wenbin",
    title = "{Energy-energy correlators in e+e- annihilation and deep inelastic scattering}",
    eprint = "2512.15896",
    archivePrefix = "arXiv",
    primaryClass = "hep-ph",
    doi = "10.1103/1lll-y39l",
    journal = "Phys. Rev. D",
    volume = "113",
    number = "7",
    pages = "074027",
    year = "2026"
}

@article{Weinberg_PhysRev.135.B1049,
  title = {Photons and Gravitons in $S$-Matrix Theory: Derivation of Charge Conservation and Equality of Gravitational and Inertial Mass},
  author = {Weinberg, Steven},
  journal = {Phys. Rev.},
  volume = {135},
  issue = {4B},
  pages = {B1049--B1056},
  numpages = {0},
  year = {1964},
  month = {Aug},
  publisher = {American Physical Society},
  doi = {10.1103/PhysRev.135.B1049},
  url = {https://link.aps.org/doi/10.1103/PhysRev.135.B1049}
}

@article{ZEUS:2023zie,
    author = "Abt, I. and others",
    collaboration = "ZEUS Collaboration",
    title = "{Measurement of jet production in deep inelastic scattering and NNLO determination of the strong coupling at ZEUS}",
    eprint = "2309.02889",
    archivePrefix = "arXiv",
    primaryClass = "hep-ex",
    reportNumber = "DESY-23-129",
    doi = "10.1140/epjc/s10052-023-12180-9",
    journal = "Eur. Phys. J. C",
    volume = "83",
    number = "11",
    pages = "1082",
    year = "2023"
}

@article{patella_2017_qed_corrections,
    author = "Patella, Agostino",
    title = "{QED Corrections to Hadronic Observables}",
    eprint = "1702.03857",
    archivePrefix = "arXiv",
    primaryClass = "hep-lat",
    reportNumber = "CERN-TH-2017-019",
    doi = "10.22323/1.256.0020",
    journal = "Proc. Sci.",
    volume = "LATTICE2016",
    pages = "020",
    year = "2017"
}

@article{autieri_2026_qT_qcdotimesqed,
    author = "Autieri, Andrea and Camarda, Stefano and Cieri, Leandro and Ferrera, Giancarlo and Sborlini, German F. R.",
    title = "{Transverse-momentum resummation at mixed QCD{\ensuremath{\otimes}}QED NNLL accuracy for Z boson production at hadron colliders}",
    eprint = "2511.07324",
    archivePrefix = "arXiv",
    primaryClass = "hep-ph",
    reportNumber = "FTUV-25-0731.8337",
    doi = "10.1007/JHEP03(2026)113",
    journal = "JHEP",
    volume = "03",
    number = "113",
    year = "2026"
}

@article{Autieri_2023_QED_QCD_qT_WZ,
   title={Combining QED and QCD transverse-momentum resummation for W and Z boson production at hadron colliders},
   volume={2023},
   ISSN={1029-8479},
   url={http://dx.doi.org/10.1007/JHEP07(2023)104},
   DOI={10.1007/jhep07(2023)104},
   number={7},
   journal={Journal of High Energy Physics},
   publisher={Springer Science and Business Media LLC},
   author={Autieri, Andrea and Cieri, Leandro and Ferrera, Giancarlo and Sborlini, German F. R.},
   year={2023},
   month={July} }

@article{Cieri_2018_QED_QCD_qT_Z,
   title={Combining QED and QCD transverse-momentum resummation for Z boson production at hadron colliders},
   volume={2018},
   ISSN={1029-8479},
   url={http://dx.doi.org/10.1007/JHEP08(2018)165},
   DOI={10.1007/jhep08(2018)165},
   number={8},
   journal={Journal of High Energy Physics},
   publisher={Springer Science and Business Media LLC},
   author={Cieri, Leandro and Ferrera, Giancarlo and Sborlini, Germán F. R.},
   year={2018},
   month={Aug} }

@article{buonocore_2024_qcdelectroweak_DY,
    author = "Buonocore, Luca and Rottoli, Luca and Torrielli, Paolo",
    title = "{Resummation of combined QCD-electroweak effects in Drell Yan lepton-pair production}",
    eprint = "2404.15112",
    archivePrefix = "arXiv",
    primaryClass = "hep-ph",
    reportNumber = "ZU-TH 21/24, CERN-TH-2024-045",
    doi = "10.1007/JHEP07(2024)193",
    journal = "JHEP",
    volume = "07",
    number = "193",
    year = "2024"
}

@article{deFlorian:2025yar,
    author = "de Florian, Daniel and Conte, Lucas Palma",
    title = "{Analytical solution for QCD $\otimes $ QED evolution}",
    eprint = "2505.03520",
    archivePrefix = "arXiv",
    primaryClass = "hep-ph",
    doi = "10.1140/epjc/s10052-026-15343-6",
    journal = "Eur. Phys. J. C",
    volume = "86",
    number = "2",
    pages = "118",
    year = "2026"
}

@article{Lepage_Brodsky_1980,
  title = {Exclusive processes in perturbative quantum chromodynamics},
  author = {Lepage, G. Peter and Brodsky, Stanley J.},
  journal = {Phys. Rev. D},
  volume = {22},
  issue = {9},
  pages = {2157--2198},
  numpages = {0},
  year = {1980},
  month = {Nov},
  publisher = {American Physical Society},
  doi = {10.1103/PhysRevD.22.2157},
  url = {https://link.aps.org/doi/10.1103/PhysRevD.22.2157}
}

@article{QW:evo,
    author = {Qiu, Jian-Wei and Watanabe, Kazuhiro},
    title = {Single inclusive hadron and jet production in lepton-hadron scattering},
    journal = {Progress of Theoretical and Experimental Physics},
    pages = {ptag170},
    year = {2026},
    month = {09},
    issn = {2050-3911},
    doi = {10.1093/ptep/ptag170},
    url = {https://doi.org/10.1093/ptep/ptag170}
}

@misc{Rubakov:2004eb,
    author = "Rubakov, V. A.",
    title = "{Lorentz-violating graviton masses: Getting around ghosts, low strong coupling scale and VDVZ discontinuity}",
    eprint = "hep-th/0407104",
    archivePrefix = "arXiv",
    reportNumber = "INR-TH-39-2004",
    month = "7",
    year = "2004"
}

@article{Rubakov:2008nh,
    author = "Rubakov, V. A. and Tinyakov, P. G.",
    title = "{Infrared-modified gravities and massive gravitons}",
    eprint = "0802.4379",
    archivePrefix = "arXiv",
    primaryClass = "hep-th",
    reportNumber = "ULB-TH-08-07",
    doi = "10.1070/PU2008v051n08ABEH006600",
    journal = "Phys. Usp.",
    volume = "51",
    pages = "759--792",
    year = "2008"
}

@article{Marsh:2015xka,
    author = "Marsh, David J. E.",
    title = "{Axion Cosmology}",
    eprint = "1510.07633",
    archivePrefix = "arXiv",
    primaryClass = "astro-ph.CO",
    reportNumber = "KCL-PH-TH-2015-50",
    doi = "10.1016/j.physrep.2016.06.005",
    journal = "Phys. Rept.",
    volume = "643",
    pages = "1--79",
    year = "2016"
}

@article{Silveira:1985rk,
    author = "Silveira, Vanda and Zee, A.",
    title = "{SCALAR PHANTOMS}",
    reportNumber = "DOE-ER-40048-13 P5",
    doi = "10.1016/0370-2693(85)90624-0",
    journal = "Phys. Lett. B",
    volume = "161",
    pages = "136--140",
    year = "1985"
}

@article{Cline:2013gha,
    author = "Cline, James M. and Kainulainen, Kimmo and Scott, Pat and Weniger, Christoph",
    title = "{Update on scalar singlet dark matter}",
    eprint = "1306.4710",
    archivePrefix = "arXiv",
    primaryClass = "hep-ph",
    doi = "10.1103/PhysRevD.88.055025",
    journal = "Phys. Rev. D",
    volume = "88",
    pages = "055025",
    year = "2013",
    note = "[Erratum: Phys.Rev.D 92, 039906 (2015)]"
}

@inproceedings{Alexander:2016aln,
    author = "Alexander, Jim and others",
    title = "{Dark Sectors 2016 Workshop: Community Report}",
    eprint = "1608.08632",
    archivePrefix = "arXiv",
    primaryClass = "hep-ph",
    reportNumber = "FERMILAB-CONF-16-421",
    month = "8",
    year = "2016"
}

@article{Langacker:2008yv,
    author = "Langacker, Paul",
    title = "{The Physics of Heavy $Z^\prime$ Gauge Bosons}",
    eprint = "0801.1345",
    archivePrefix = "arXiv",
    primaryClass = "hep-ph",
    doi = "10.1103/RevModPhys.81.1199",
    journal = "Rev. Mod. Phys.",
    volume = "81",
    pages = "1199--1228",
    year = "2009"
}

@article{Han:1998sg,
    author = "Han, Tao and Lykken, Joseph D. and Zhang, Ren-Jie",
    title = "{On Kaluza-Klein states from large extra dimensions}",
    eprint = "hep-ph/9811350",
    archivePrefix = "arXiv",
    reportNumber = "MADPH-98-1092, FERMILAB-PUB-98-364",
    doi = "10.1103/PhysRevD.59.105006",
    journal = "Phys. Rev. D",
    volume = "59",
    pages = "105006",
    year = "1999"
}

@article{DAmico:2011eto,
    author = "D'Amico, G. and de Rham, C. and Dubovsky, S. and Gabadadze, G. and Pirtskhalava, D. and Tolley, A. J.",
    title = "{Massive Cosmologies}",
    eprint = "1108.5231",
    archivePrefix = "arXiv",
    primaryClass = "hep-th",
    reportNumber = "NYU-TH-08-28-11",
    doi = "10.1103/PhysRevD.84.124046",
    journal = "Phys. Rev. D",
    volume = "84",
    pages = "124046",
    year = "2011"
}

@article{Alberte:2019xfh,
    author = "Alberte, Lasma and de Rham, Claudia and Momeni, Arshia and Rumbutis, Justinas and Tolley, Andrew J.",
    title = "{Positivity Constraints on Interacting Spin-2 Fields}",
    eprint = "1910.11799",
    archivePrefix = "arXiv",
    primaryClass = "hep-th",
    reportNumber = "Imperial/TP/2019/LA/03",
    doi = "10.1007/JHEP03(2020)097",
    ISSN={1029-8479},
    url={http://dx.doi.org/10.1007/JHEP03(2020)097},
    journal = "JHEP",
    number={3},
    volume={2020},
    pages = "097",
    year = "2020"
}

@article{Vainshtein:1972sx,
    author = "Vainshtein, A. I.",
    title = "{To the problem of nonvanishing gravitation mass}",
    doi = "10.1016/0370-2693(72)90147-5",
    journal = "Phys. Lett. B",
    volume = "39",
    pages = "393--394",
    year = "1972"
}

@article{Babichev:2013usa,
    author = "Babichev, Eugeny and Deffayet, C{\'e}dric",
    title = "{An introduction to the Vainshtein mechanism}",
    eprint = "1304.7240",
    archivePrefix = "arXiv",
    primaryClass = "gr-qc",
    reportNumber = "LPT-ORSAY-13-45",
    doi = "10.1088/0264-9381/30/18/184001",
    journal = "Class. Quant. Grav.",
    volume = "30",
    pages = "184001",
    year = "2013"
}

@article{Hou:2019efy,
    author = "Hou, Tie-Jiun and others",
    title = "{New CTEQ global analysis of quantum chromodynamics with high-precision data from the LHC}",
    eprint = "1912.10053",
    archivePrefix = "arXiv",
    primaryClass = "hep-ph",
    reportNumber = "MSUHEP-19-025, PITT-PACC-1911, SMU-HEP-19-03",
    doi = "10.1103/PhysRevD.103.014013",
    journal = "Phys. Rev. D",
    volume = "103",
    number = "1",
    pages = "014013",
    year = "2021"
}
\end{document}